\documentclass[fleqn,usenatbib]{mnras}

\usepackage{newtxtext,newtxmath}

\usepackage[T1]{fontenc}

\DeclareRobustCommand{\VAN}[3]{#2}
\let\VANthebibliography\thebibliography
\def\thebibliography{\DeclareRobustCommand{\VAN}[3]{##3}\VANthebibliography}

\usepackage{graphicx}	
\usepackage{amsmath}	
\usepackage{caption}
\usepackage{float}
\usepackage{gensymb}
\usepackage{subfig}

\title[QPO phenomenon in MAXI J1535-571 ]{On the energy dependence of the QPO phenomenon in the black hole system MAXI J1535-571}

\author[Garg et al.]{
Akash Garg,$^{1}$\thanks{E-mail: akashgarg\_16@yahoo.co.in}
Ranjeev Misra,$^{2}$
Somasri Sen$^{1}$
\\
$^{1}$Department of Physics, Jamia Millia Islamia, Jamia Nagar, New Delhi-110025, India\\
$^{2}$Inter-University Centre for Astronomy and Astrophysics, Ganeshkhind, Pune-411007, India\\
}

\date{Accepted XXX. Received YYY; in original form ZZZ}

\pubyear{2021}

\begin{document}
\label{firstpage}
\pagerange{\pageref{firstpage}--\pageref{lastpage}}
\maketitle

\begin{abstract}
Previous analysis of \textit{AstroSat} observations of the black hole system MAXI J1535-571, have revealed the presence of a strong Quasi-Periodic Oscillation (QPO) whose frequency is correlated with the high energy spectral index. Here, we fit the spectra as emitted from a truncated disc with an inner hot corona, study the QPO frequency dependence on other spectral parameters and model the energy dependent r.m.s and time-lag of the QPO to identify the physical spectral parameters whose variation are responsible for the QPO. The QPO frequency is found to also correlate with the scattering fraction (i.e. the fraction of the soft photons Comptonized) and its dependence on the accretion rate and inner disc radii is consistent with it being the dynamical frequency. The time-lag between the hard and soft photons is negative for QPO frequency $> 2.2$ Hz and is positive for lesser values, making this the second black hole system to show this behaviour after GRS 1915+105. Modelling the energy dependent time-lag and r.m.s requires correlated variation of the accretion rate, inner disc radii and the coronal heating rate, with the latter having a time-lag compared to the other two for QPO frequencies less than $< 2.2$ Hz and  which changes sign (i.e. the coronal heating variation precedes the accretion rate one) for higher values. The implications of the results are discussed.

\end{abstract}




\section{Introduction}

X-ray binaries form a class of astronomical objects that contain a pair of massive bodies orbiting around their center of mass. Black hole X-ray binaries (hereafter, BHXBs) in particular, host a black hole and a non-collapsed star. During its evolution, the normal star may swell up and overflow its tidal lobe resulting in matter transfer through a common Lagrangian point (Roche lobe overflow) or it may also eject gaseous matter in the form of a stellar wind (Stellar wind accretion). The matter is captured by strong gravity of the black hole. As the matter undergoes in-fall, the angular momentum possessed by it is transferred outwards leading to the formation of planar structures around the black hole, known as accretion discs. Such an accretion phenomenon in the inner regions emit profusely in the X-rays and other parts of electromagnetic radiation~(\cite{2002apa}). 

Most BHXBs are transients and are known observationally to either stay dormant (low flux/quiescence state) or to undergo outburst (high flux/active state) during their evolution. Such a slow and long term systematic variation is also accompanied by rapid variations in the X-ray emission. While long term variability can be inferred through visual inspection of the lightcurves, short term variations need to be quantified using Power Density Spectrum (PDS). The PDS is defined as being proportional to the square of the Fourier amplitudes, obtained by taking the Fourier transform of the light curve (\cite{1989ASIC..262...27V}). The computed PDS for different BHXBs are often characterised by broad features (sometimes referred to as broadband noise) along with localised peaks known as Quasi-periodic Oscillations (QPOs). \cite{1997MNRAS.292..679L} showed that the broadband variability in the PDS could be due to fluctuations in the viscosity parameter at different radii of the Shakura-Sunyaev geometrically thin disc~(\cite{1973A&A....24..337S}). These fluctuations induce variations in the accretion rate which propagate inwards to smaller radii of the disc, resulting in the variation of the observed flux. Such a propagating fluctuations model may also explain the hard lags (i.e. high energy photons arriving later than the low energy ones) observed in these sources.

QPOs have been detected for several BHXBs over the past three decades. Broadly they are categorized as Low Frequency QPOs (LFQPOs, $\le$ 30 Hz) and High Frequency QPOs (HFQPOs, $\ge$ 60 Hz). The QPOs are generally fitted with a Lorentzian function making it possible to measure their properties like centroid frequency $\nu$, strength and quality factor $\nu/\Delta \nu$. This leads to the further classification of LFQPOs as Type A, B, and C QPOs~(see review by \cite{2019NewAR..8501524I}). QPOs are thought to originate from the inner regions of the accretion flow and hence provide the possibility to study extreme gravity near the black hole. Despite multiple detection of type-C LFQPOs, the physics of their origin is not yet clear. There exist several models to explain type-C QPOs (sometimes type-B QPOs as well) which are based on the idea of a truncated disc geometry where a geometrically thin and optically thick accretion disc truncates at a radius greater than Innermost Stable Circular Orbit (ISCO) around the black hole and the region inside it is filled with hot thermal plasma. Furthermore, geometrical effects and thermal instabilities in the accretion disc are considered to explain observed variability. For instance, Relativistic precession model~ (\cite{2009MNRAS.397L.101I,2018MNRAS.473..431M,2020ApJ...903..121R}), Accretion-ejection instability~(\cite{1999A&A...349.1003T}), propagating oscillatory shock model~(\cite{1996ApJ...457..805M,2000ApJ...531L..41C}), transition layer model~(\cite{2004ApJ...612..988T}) and many more had been proposed to identify the QPO frequency with their respective theoretical predictions.

During a typical outburst, a BHXB passes through distinct spectral/timing states as seen in the  hardness-Intensity diagram (HID) which typically has a "q" shaped pattern (\cite{2016AN....337..398M,2016ASSL..440...61B}). It should be noted that the precise identification of the state requires timing information (\cite{2016ASSL..440...61B}). The main evolution which goes like - Low hard state (LHS)-Hard-intermediate state (HIMS)-Soft Intermediate state (SIMS)-HSS (High soft state) shows systematic changes in the variability amplitude as well as the appearance/disappearance of type-C LFQPOs in different states~(see \cite{2011BASI...39..409B}). This indicates a connection between QPO frequency and spectral nature of the source and provides an alternative method to understand the origin of QPOs. This has been confirmed repeatedly by showing strong correlations of QPO centroid frequency with both disc and power-law parameters like inner disc radius and photon index respectively~(see \cite{2003A&A...397..729V} and references therein). The observed QPOs also possess energy-dependent properties like fractional r.m.s and time-lag, giving crucial information about the underlying dynamic process in the accretion disc~(\cite{2000ApJ...541..883R,2001ApJ...556..515M,2013MNRAS.436.2334P}). \cite{2013ApJ...779...71M} and \cite{2016MNRAS.457.2999M} showed how a simple phenomenological model which considers variations in spectral parameters can explain the QPO's energy dependent behaviour.

The aforementioned results utilized data from the Rossi X-ray timing explorer (\textit{RXTE}). Although \textit{RXTE} discovered multiple QPOs in BHXB, its effective energy range of 3-20 keV is insufficient to determine the nature of QPOs at lower and higher energies. Large X-ray proportional counter (\textbf{LAXPC})~(\cite{2016SPIE.9905E..1DY,2017JApA...38...30A}) onboard \textit{AstroSat} captures BHXB in the broader energy band of 3-50 keV along with low energy coverage (1-7 keV) provided by Soft X-ray telescope (\textbf{SXT})~(\cite{2017JApA...38...29S}).  \cite{2019MNRAS.486.2964M}, \cite{2019ApJ...887..101J}, and \cite{2020ApJ...889L..17M} illustrated the spectro-timing capabilities of \textit{AstroSat} by modelling energy-dependent r.m.s and lag in both persistent (Cygnus X-1) and transient systems (MAXI J1820+070, SWIFT J1658.2–4242) through a one-zone stochastic propagation model. These work use a specific spectral model like thermal comptonization to predict the timing properties.

In \cite{2020MNRAS.498.2757G}, we outlined a generic technique where radiative components of time-averaged photon spectrum can be varied to model the observed timing behavior. As an example, we applied it to energy dependent features of LFQPOs detected in BHXB GRS 1915+105 using \textit{AstroSat}/\textbf{LAXPC} data. We considered a simple but often seen case where photon spectrum can be modelled as emission from a truncated accretion disc surrounding a hot thermal plasma. Though we could compute the magnitude of variations in different spectral parameters that can produce the observed fractional r.m.s and lag, it was difficult to infer evolution of timing model parameters and find any correlations owing to smaller set of observations.

In the present work, we consider data of a black hole transient MAXI J1535-571 as observed by \textbf{LAXPC} and \textbf{SXT} over a period of five days. \cite{2019MNRAS.487..928S} detected type-C QPOs in the frequency range 1.85 Hz-2.88 Hz in the hard-intermediate state of the source. \cite{2019MNRAS.487..928S} and \cite{2019MNRAS.487.4221S} both performed broadband spectral modelling and estimated the black hole mass to be 5.14-7.83 $M_{\odot}$ and 10.39 $M_{\odot}$ respectively. However, the absence of dynamical mass measurement implies that the binary parameters, such as the mass of the black hole, the inclination angle are not certain. \cite{2019MNRAS.488..720B} found 1.7-3.0 Hz QPO frequency to be tightly correlated with power law spectral index. In this work, we apply the generic technique as outlined in \cite{2020MNRAS.498.2757G} to understand the QPO phenomenon. 

In the next Section, we present the spectro-timing analysis, in Section 3 we model the timing properties of observed QPOs and present conclusions and discussion in the last Section.

\section{Spectral and timing analysis}

\textit{AstroSat} observed MAXI J1535-571 through target of opportunity (ToO) from September 12 to 17, 2017. The total stare time of the observation was 400 ksecs. We obtained level 1 \textbf{LAXPC} and \textbf{SXT} data from the ISSDC data archive and reduced it into level 2 event files using latest \textbf{LAXPC}\footnote{\url{http://astrosat-ssc.iucaa.in/laxpcData}} and \textbf{SXT}\footnote{\url{http://astrosat-ssc.iucaa.in/sxtData}} software packages. \cite{2019MNRAS.488..720B} had divided the whole of data into 66 segments with similar exposures (Refer to Table 1 in \cite{2019MNRAS.488..720B} ). We followed the same division to carry out spectral and timing analysis for each of the segment using data from detectors \textbf{LAXPC10}, \textbf{LAXPC20} and \textbf{SXT}.

\begin{figure}
\includegraphics[width=0.52\textwidth,height=6cm]{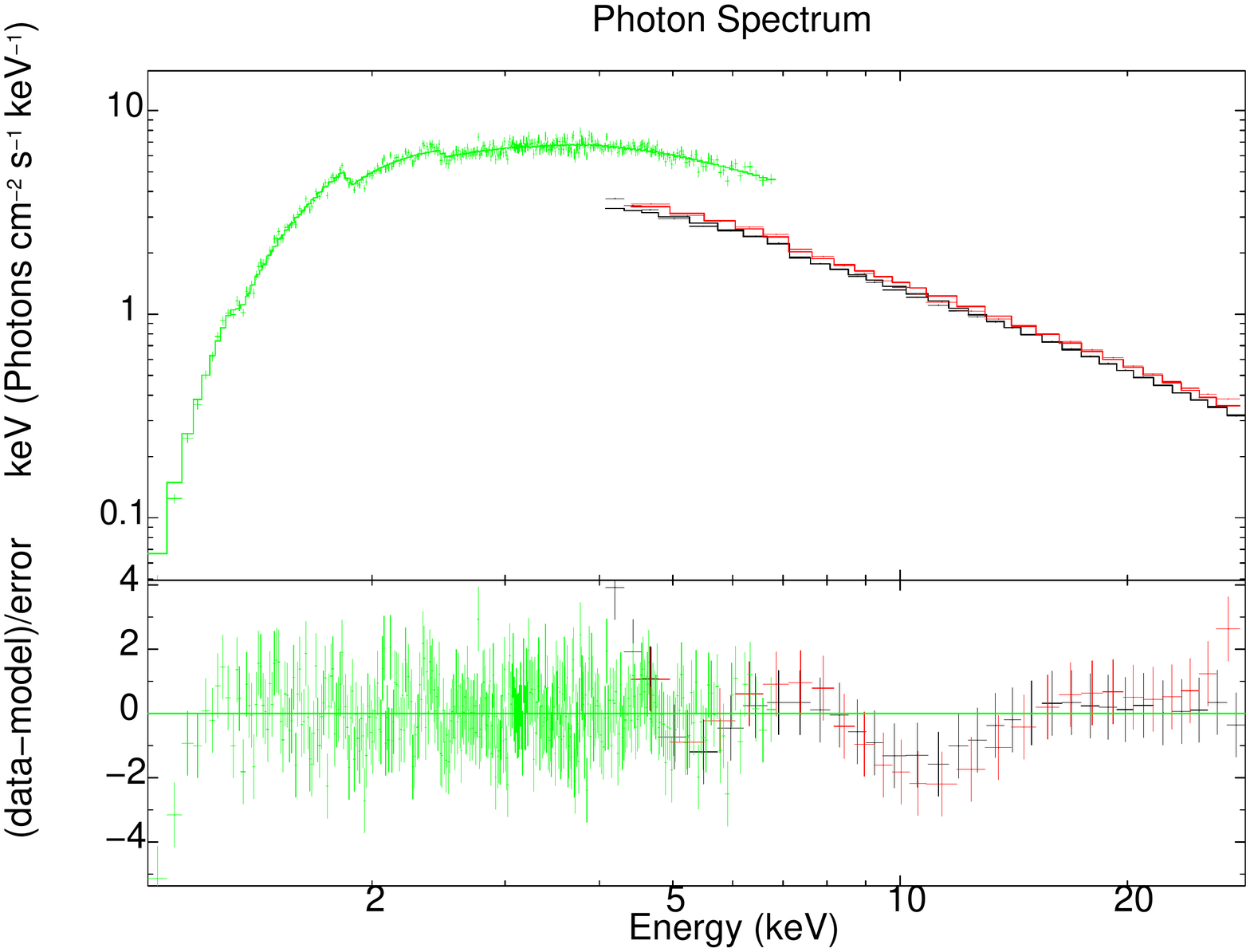}
\caption{Photon spectrum in 1-30 keV as observed by \textbf{LAXPC10} (Black), \textbf{LAXPC20} (Red) and \textbf{SXT} (Green). Top panel shows the fitting of data with XSPEC model: \textit{constant*TBpcf*TBabs*ThComp*diskbb}. Bottom panel shows residuals.}
\label{fig:spectrum}
\end{figure}

\cite{2019MNRAS.487..928S} found that the time-averaged broadband photon spectrum in the energy range 0.7-80 keV can be fitted with XSPEC component \textit{nthcomp} - a continuum thermal comptonization model~(\cite{1996MNRAS.283..193Z,1999MNRAS.309..561Z}) better than the combination of \textit{diskbb}~(\cite{1984PASJ...36..741M}) and \textit{powerlaw}. Although \cite{2019MNRAS.488..720B} used the same model components, they restricted the analysis to 1-30 keV, because of the background uncertainties beyond 30 keV for \textbf{LAXPC10} and \textbf{20} detectors and the uncertain energy response for \textbf{SXT} below 1 keV. Though these two works couldn't find significant iron-line emission for a particular segment, \cite{2019MNRAS.487.4221S} found its presence and thereby included reflection models along with \textit{diskbb} and \textit{nthcomp} to constrain blackhole mass and spin. 
\begin{figure*}
\includegraphics[width=\textwidth,height=11cm]{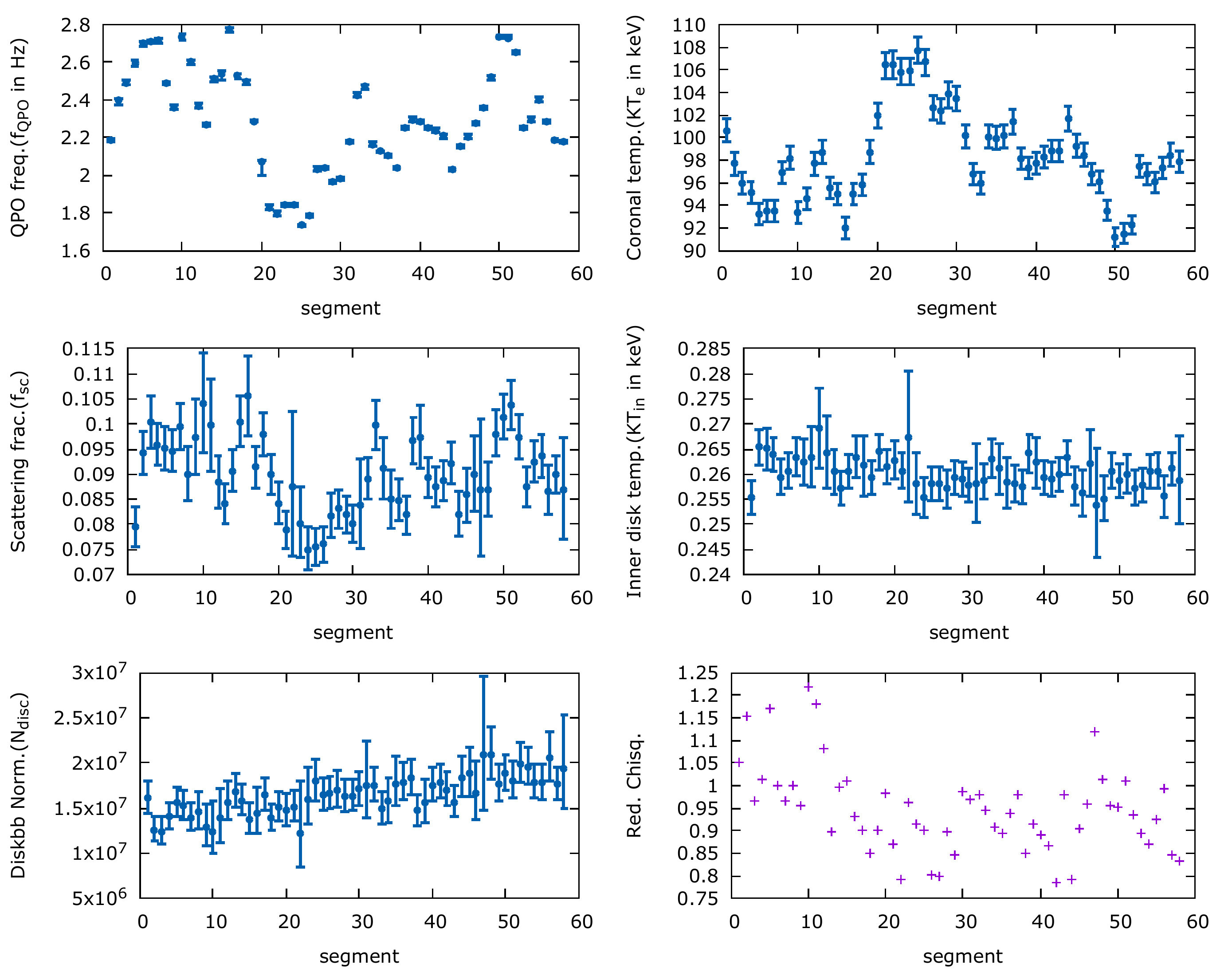}
\caption{Top left panel shows the behavior of centroid frequency of QPO. Rest of the panels show the evolution of spectral parameters and reduced chi-square over different segments. It is to noted that for some of the segments, \textbf{SXT} exposure is zero and hence we left them during spectral analysis.}
\label{fig:spectrumpar}
\end{figure*}

\begin{figure*}
\includegraphics[width=0.9\textwidth,height=9cm]{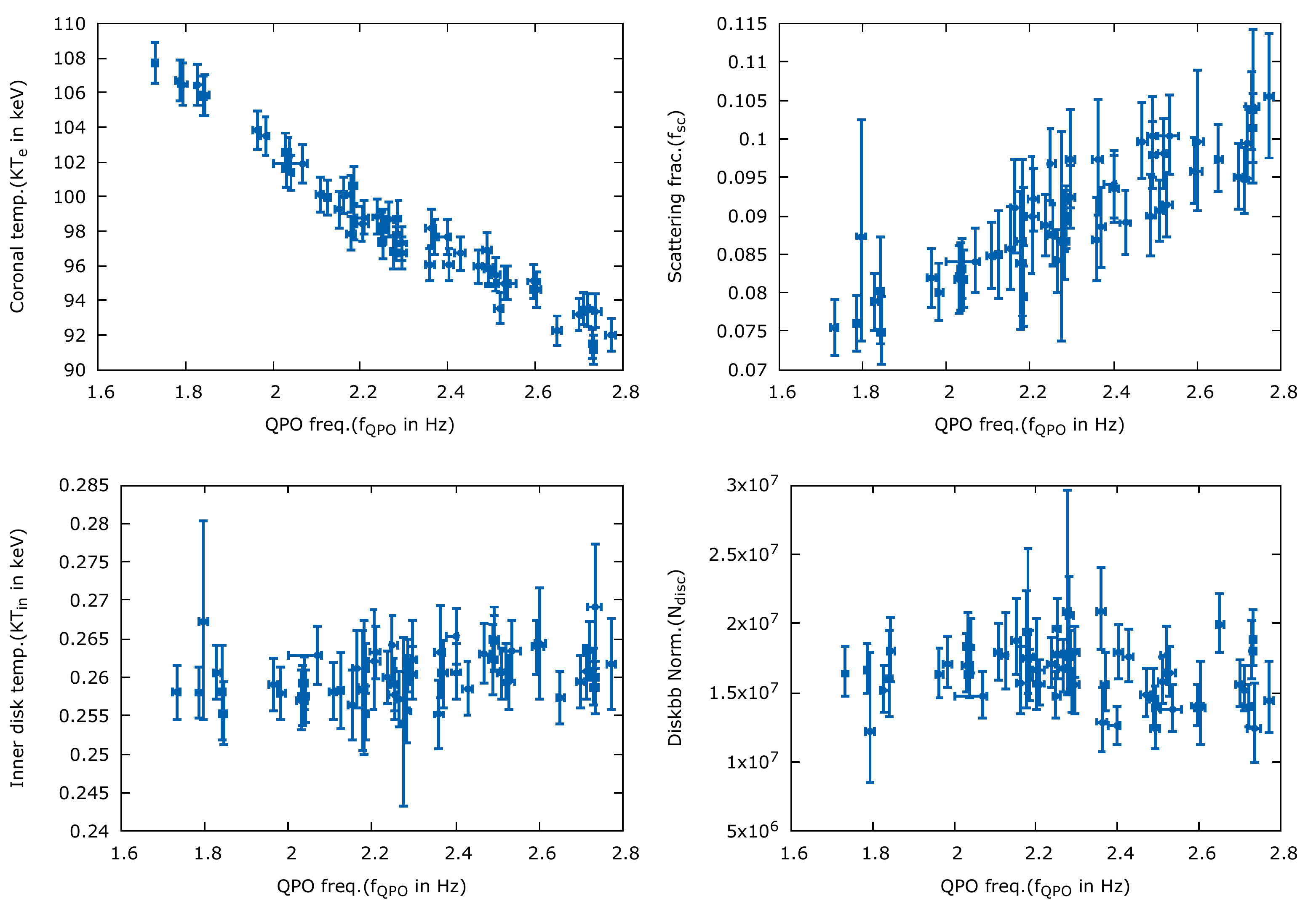}
\caption{Correlation of spectral parameters with QPO frequency.}
\label{fig:spectrumparf}
\end{figure*}

\begin{figure}
\includegraphics[width=0.48\textwidth,height=10cm]{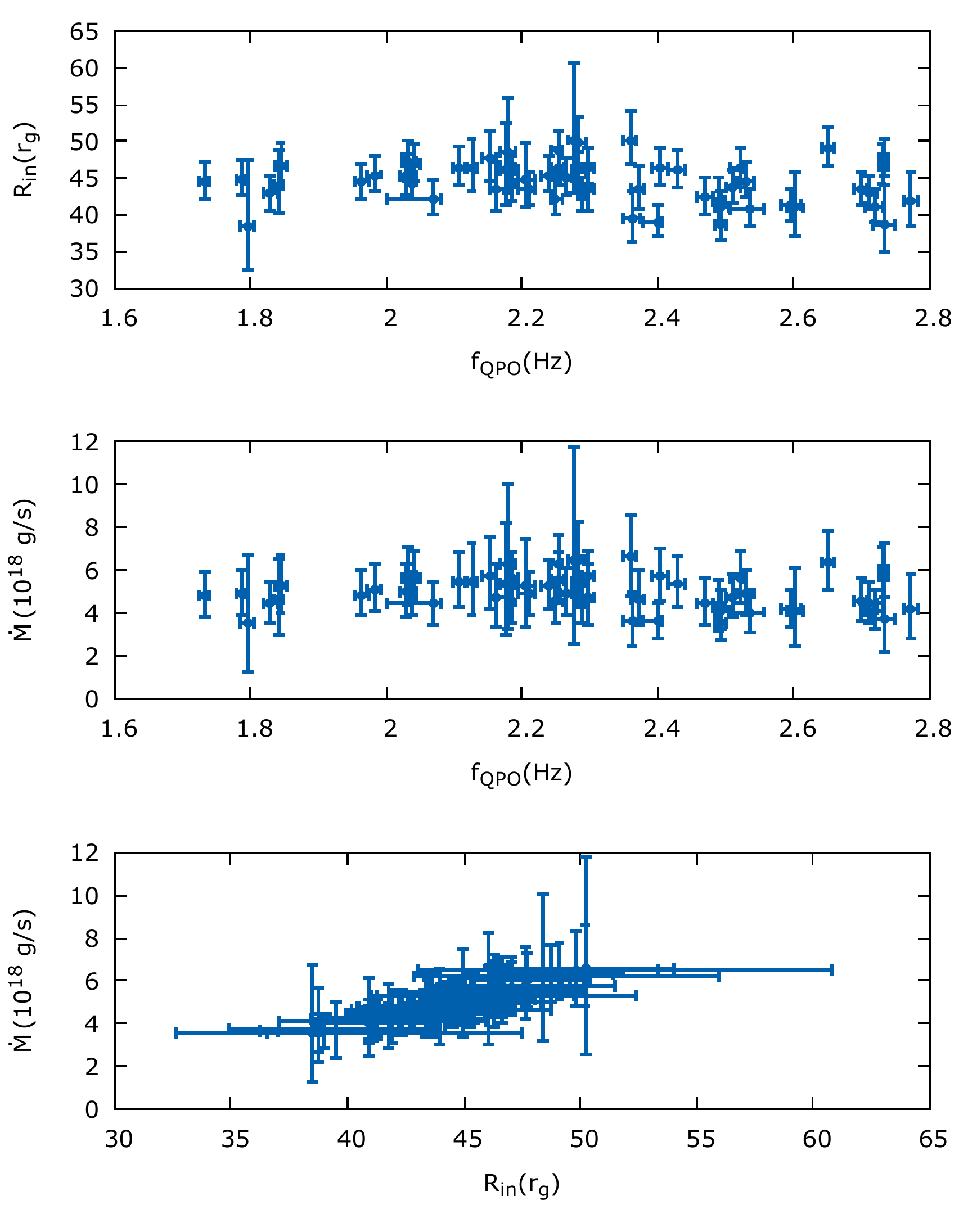}
\caption{Top and middle panel show correlation of inner disc radii and accretion rate with QPO frequency respectively. Bottom panel shows the relation between inner disc radii and accretion rate.}
\label{fig:spectrumrin}
\end{figure}

\begin{figure}
\includegraphics[width=0.48\textwidth,height=5cm]{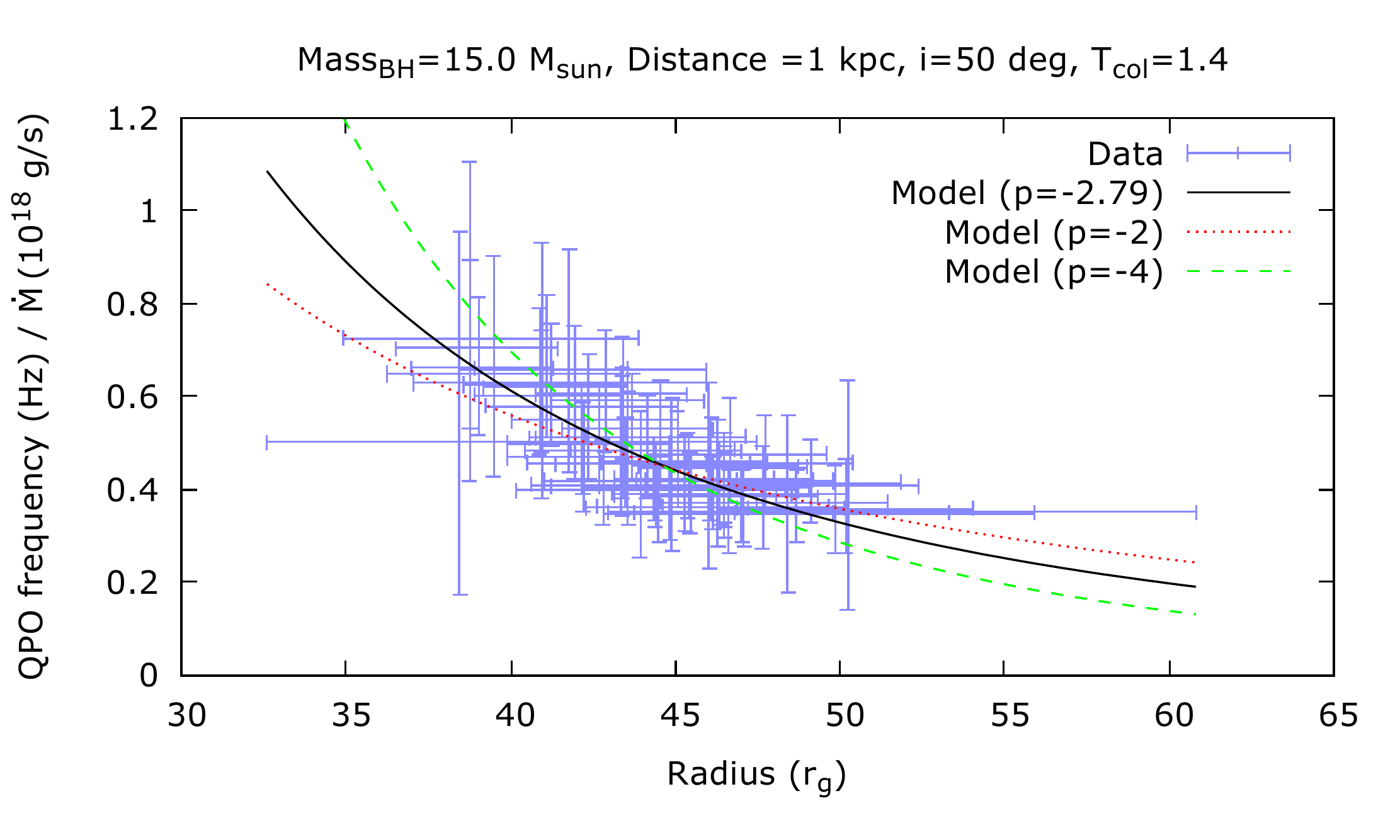}
\caption{Shows the correlation between $f_{QPO}/\Dot{M}$ and $R_{in}$. The solid line is the best fit corresponding to the function $f_{QPO}/\Dot{M}~\propto~R_{in}^{p}$ where $p = -2.79\pm0.25$. The dotted and dashed lines are for the same function with p = -2 and p = -4 respectively.}
\label{fig:qpomrin}
\end{figure}

In the present work, we first modelled the spectrum (1-30 keV) for a typical segment using \textit{diskbb} and \textit{ThComp} to account for emissions from outer thermal disc and inner region of coronal plasma. \textit{ThComp} is a convolution model and works as a substitution for \textit{nthcomp} to model the thermal spectrum emitted by a spherical source of electrons~(\cite{2020MNRAS.492.5234Z}). \textit{ThComp} provides the flexibility of having as a parameter the scattering fraction $f_{sc}$, which is the fraction of the soft seed photons that are Comptonized by the hot medium. The absorption column density $N_{H}$ is accounted for using XSPEC component \textit{Tbabs}~(\cite{2000ApJ...542..914W}). The spectral fitting gave a reduced chi-square to be 1.45 for 402 degrees of freedom.
The addition of a partial covering fraction absorption model \textit{Tbpcf}, significantly improved the fitting with the reduced $\chi^2$  decreasing to 1.04. Figure~\ref{fig:spectrum} shows the source's unfolded spectrum for this particular segment. Such a partial covering feature was also used by \cite{2019ApJ...887..101J} to model the spectra of the black hole system SWIFT J1658.2–4242. Furthermore, we found that for all the segments the spectral fitting seemed to require dual absorption components, one of them having partial covering. The two $N_H$ and the partial covering fraction (pcf) didn't vary much and hence we fixed these parameters to their average value i.e. $\sim$ $4.08\times10^{22} cm^{-2}$, $3.27\times10^{22} cm^{-2}$ and 0.82 respectively.

In thermal Comptonizaton the high energy  spectral index $\Gamma$ is determined using two physical parameters - the electron temperature $KT_e$ and the optical depth $\tau$ of the coronal cloud. The XSPEC model \textit{ThComp} provides the option of using the Thomson optical depth ($\tau$) instead of the spectral index ($\Gamma$) as a fitting parameter. As we describe later, this is more convenient in predicting the energy dependent time-lag and r.m.s for the system. Since we couldn't constrain the coronal temperature, it was fixed at 100 keV and the best fit optical depth for the segment under consideration came out to be $\sim$ 0.65. We chose to keep optical depth fixed to 0.65 for all segments and allowed for the electron temperature $kT_e$ to be a free parameter. It would have also been possible to do the reverse, i.e. fix the temperature to 100 keV for all segments and allowed for the optical depth to vary, however it is perhaps more physical to allow temperature variation responding to a change in the heating rate.

The power density spectrum for each of the segment was obtained using \textbf{LAXPC} analysis tool laxpc$\_$find$\_$freqlag\footnote{\url{http://www.tifr.res.in/~astrosat_laxpc/LaxpcSoft.html}} in a broad frequency range 0.1-100 Hz and QPOs in the range $\sim$ 1.7-2.8 Hz were identified as reported by \cite{2019MNRAS.488..720B}. The QPO frequencies, the best fit spectral parameters as well as the reduced $\chi^2$ as a function of segment number (i.e. time) are shown in different panels of Figure~\ref{fig:spectrumpar}. Among the spectral parameters, the most prominent variation is seen in the electron temperature $kT_e$, while the fractional scattering also varies. Figure~\ref{fig:spectrumparf} shows the variation of the spectral parameters with QPO frequency, where a clear anti-correlation between the electron temperature and QPO frequency is seen, which is expected given the correlation reported by \cite{2019MNRAS.488..720B} between index $\Gamma$ and the QPO frequency. However, there is also a correlation between the scattering fraction and QPO frequency. 

\begin{figure}

\centering
\includegraphics[width=0.45\textwidth,height=9cm]{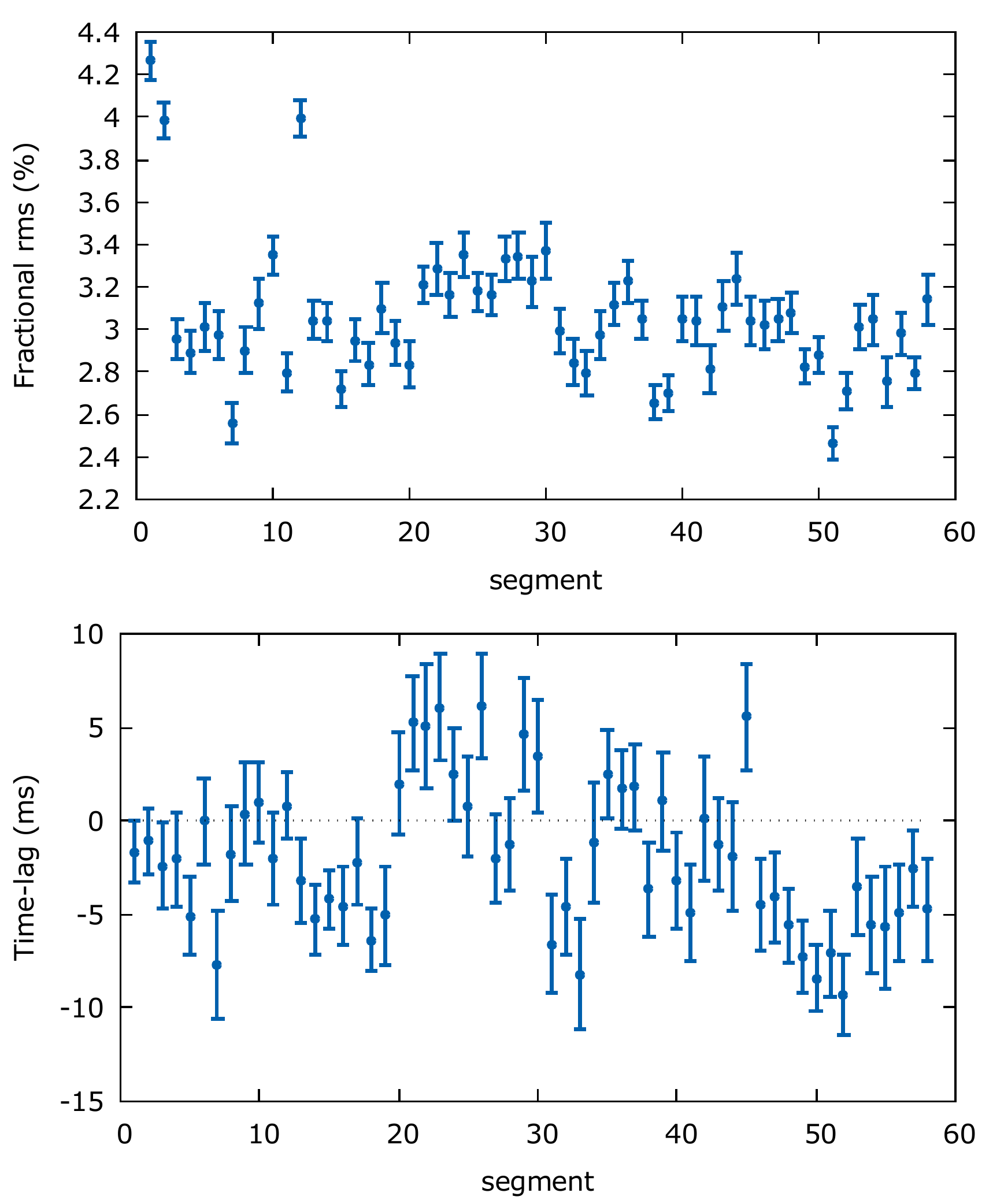}

\caption{Top panel shows the evolution of fractional r.m.s in the energy range 4-30 keV whereas bottom panel shows the evolution of time-lag of 12-15 keV with respect to 4-5.38 keV.}
\label{fig:rms_lag}
\end{figure}

\begin{figure}

\centering
\includegraphics[width=0.47\textwidth,height=4.8cm]{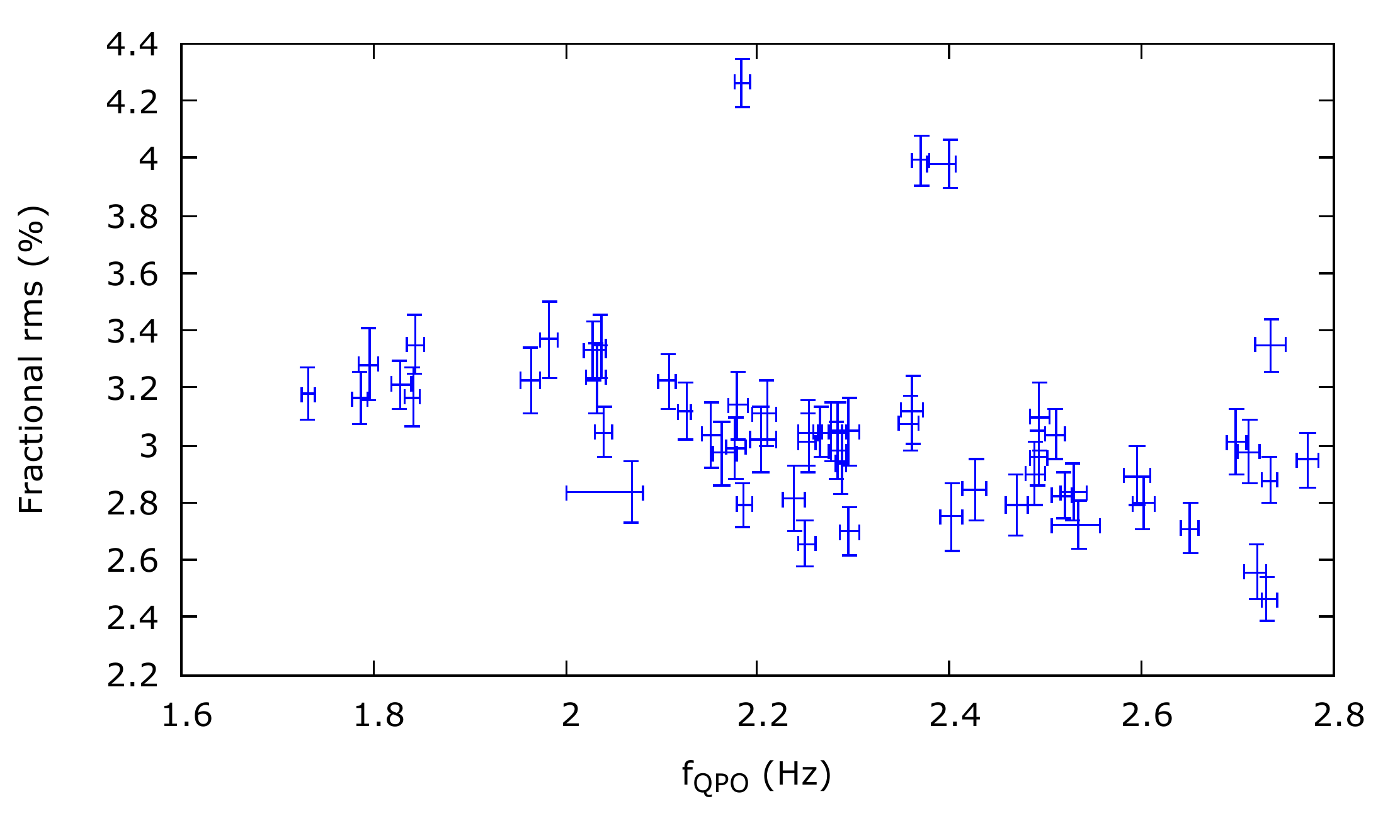}\hfill
\includegraphics[width=0.47\textwidth,height=4.8cm]{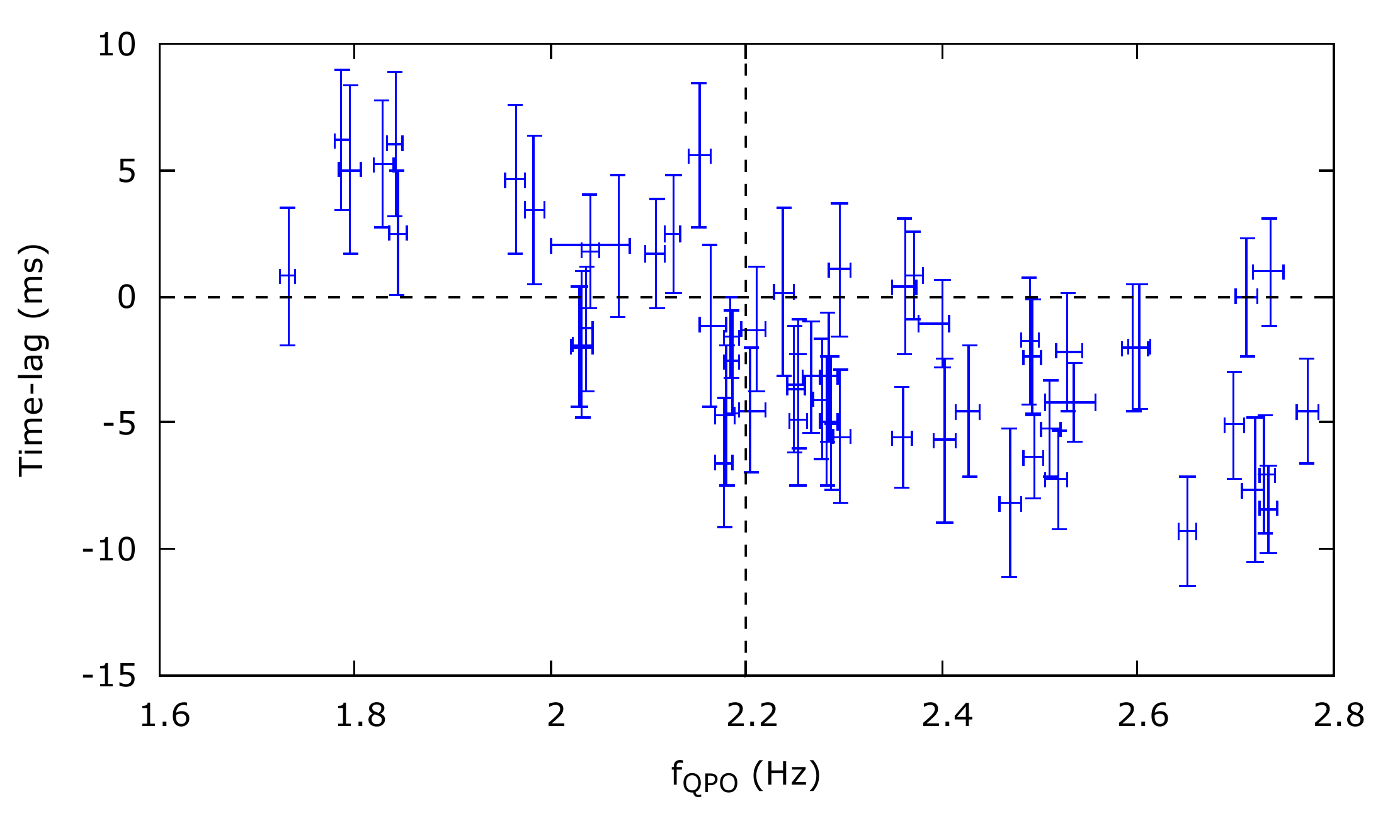}

\caption{Top and bottom panel show correlation of fractional r.m.s in 4-30 keV and time-lag of 12-15 keV with respect to 4-5.38 keV with frequency respectively.}
\label{fig:rlf}
\end{figure}

The QPO frequency does not seem to depend on the disc spectral parameters (i.e. the normalization and inner disc temperature). Moreover, the disc spectral parameters do not vary significantly in contrast to the QPO frequency and coronal spectral parameters. To obtain further insight, we translate the best fit disc spectral parameters to quantities related to the accretion rate and inner disc radius. Specifically, the normalization of the disc component, $N_{disc} = (R_{in}/D_{10})^2~cos~\theta$, where $D_{10}$ is the distance to the source in units of 10 kpc and $\theta$ is the inclination angle and the inner disc temperature (\cite{1984PASJ...36..741M}), $T_{in} = (\frac{3GM\Dot{M}}{8\pi R_{in}^3 \sigma})^{1/4}$, where M and $\sigma$ are the mass of the black hole and Stefan-Boltzmann constant respectively (\cite{2002apa}). We estimated $R_{in}$ and $\Dot{M}$ using the above relations and fiduciary values for the  distance of 1 kpc, blackhole mass of 15 $M_{\odot}$, inclination angle of $50^\degree$ and color factor of 1.4.  We note that changes in these assumed values would lead to scaling of the inferred accretion rate and inner disc radii. The top two panels of Figure~\ref{fig:spectrumrin} shows the correlation of $R_{in}$ and $\Dot{M}$ with QPO frequency which seemingly again shows the absence of any significant correlation between the QPO frequency and disc parameters. However, there is a positive correlation between $R_{in}$ and $\Dot{M}$ as can be seen in the bottom panel where $\dot M \propto R_{in}^{2.4}$. For the QPO and spectral observations of the black hole system, GRS 1915+105, \cite{2020ApJ...889L..36M} and \cite{2021ApJ...909...63L} have shown that the QPO frequency divided by the accretion rate has a functional dependence on the inner disc radius as predicted in General Relativity if the frequency is identified as the dynamical one (i.e. related to the sound crossing time-scale). For the data set considered here, Figure~\ref{fig:qpomrin} shows that there indeed exists a correlation between QPO frequency($f_{QPO}$)/accretion rate($\Dot{M}$) and inner disc radii which can be fitted using the functional form, $f_{QPO}/\Dot{M}~\propto~R_{in}^p$ with  $p = -2.79\pm0.25$. Note that the above correlation (and in particular the value of $p$) is independent of the assumed mass, distance, inclination angle and color factor. This becomes obvious if we recast the function $f_{QPO}/\Dot{M}~\propto~R_{in}^p$ in terms of $f_{QPO}$, $KT_{in}$ and $N_{disc}$ and by noting that although the constant of the proportionality may depend on the assumed values, the form of the  proportionality itself is independent. However, we emphasize that unlike the analysis undertaken on GRS 1915+105 by \cite{2020ApJ...889L..36M} and \cite{2021ApJ...909...63L}, the spectra of MAXI J1535-571 here is dominated by the Comptonized component and the disc component is relatively weak. We defer a discussion on these results to the last section.

We computed fractional r.m.s and time-lags for each of the segment at QPO frequency using \textbf{LAXPC} subroutine laxpc$\_$find$\_$freqlag. For a particular energy band, the subroutine calculates the fractional r.m.s for a frequency 'f' by integrating the power spectrum over the frequency resolution $\Delta$f and then taking its square root. In principal $\Delta$f is usually equal to (or half of) Full Width Half Maximum (FWHM) of QPO frequency. Additionally, for the same inputs f and $\Delta$f, the subroutine computes the phase lag of cross-spectrum of two light curves in different energy bins which is divided by $2\pi$f to obtain time-lags~(\cite{1999ApJ...510..874N}). 

First we calculated fractional r.m.s in the energy range 4-30 keV and time-lag of 12-15 keV photons with respect to a reference energy band of 4-5.38 keV for each of the QPO frequency\footnote{The energy bands were chosen such that the errors on the time-lags were minimized. Using different energy bands gives qualitatively similar results but with larger uncertainty in the time-lags.}. It can be seen from Figure~\ref{fig:rms_lag} that the time-lag is alternating between positive and negative sign. It is more evident in Figure~\ref{fig:rlf} where we plotted time-lag vs QPO frequency. The time-lag switches sign from positive to negative around $\sim$2.2 Hz. Such changes in sign at $\sim$2.2 Hz were also found by \cite{2000ApJ...541..883R,2013MNRAS.436.2334P} and \cite{2020MNRAS.494.1375Z} for low-frequency QPOs in GRS 1915+105. In addition to this, \cite{2020MNRAS.494.1375Z} showed non-monotonic behavior of fractional r.m.s where it first increases and then decrease with QPO frequency, reaching a maximum at $\sim$2 Hz. Though there is a broad scatter but yet Figure~\ref{fig:rlf} also shows the region where r.m.s decreases with frequency. Since there are lesser points below 2 Hz, it is difficult to confirm the maxima and the possible region of increase in r.m.s. Following this we also generated the fractional r.m.s and time-lag in different energy bands with respect to the same reference energy range of 4-5.38 keV. We discuss in next section of how we model this energy-dependence with the numerical scheme mentioned in \cite{2020MNRAS.498.2757G}.

\section{Modelling the energy-dependent fractional r.m.s and time-lags}

\cite{2020MNRAS.498.2757G} described a comprehensive technique to model the timing properties based on the spectral information of the source. The idea put forward was to first model the time-averaged energy spectrum with appropriate spectral components (i.e. XSPEC spectral models). Next one translates the spectral parameters into physical ones whose coordinated variation may describe the observed fractional r.m.s and time-lag. Then, one numerically computes the complex linear response of the spectrum, $\Delta F(E)$, to complex small variations of the physical parameters at a given frequency, which in turn provides predictions for the energy dependent fractional r.m.s ($(1/\sqrt{2})|\Delta F(E)|/F(E)$) and time-lag (argument of ${\Delta F(E_{ref})}^*\Delta F(E)$) to be compared with observations.

\begin{figure*}

\centering
\includegraphics[width=0.48\textwidth,height=4.8cm]{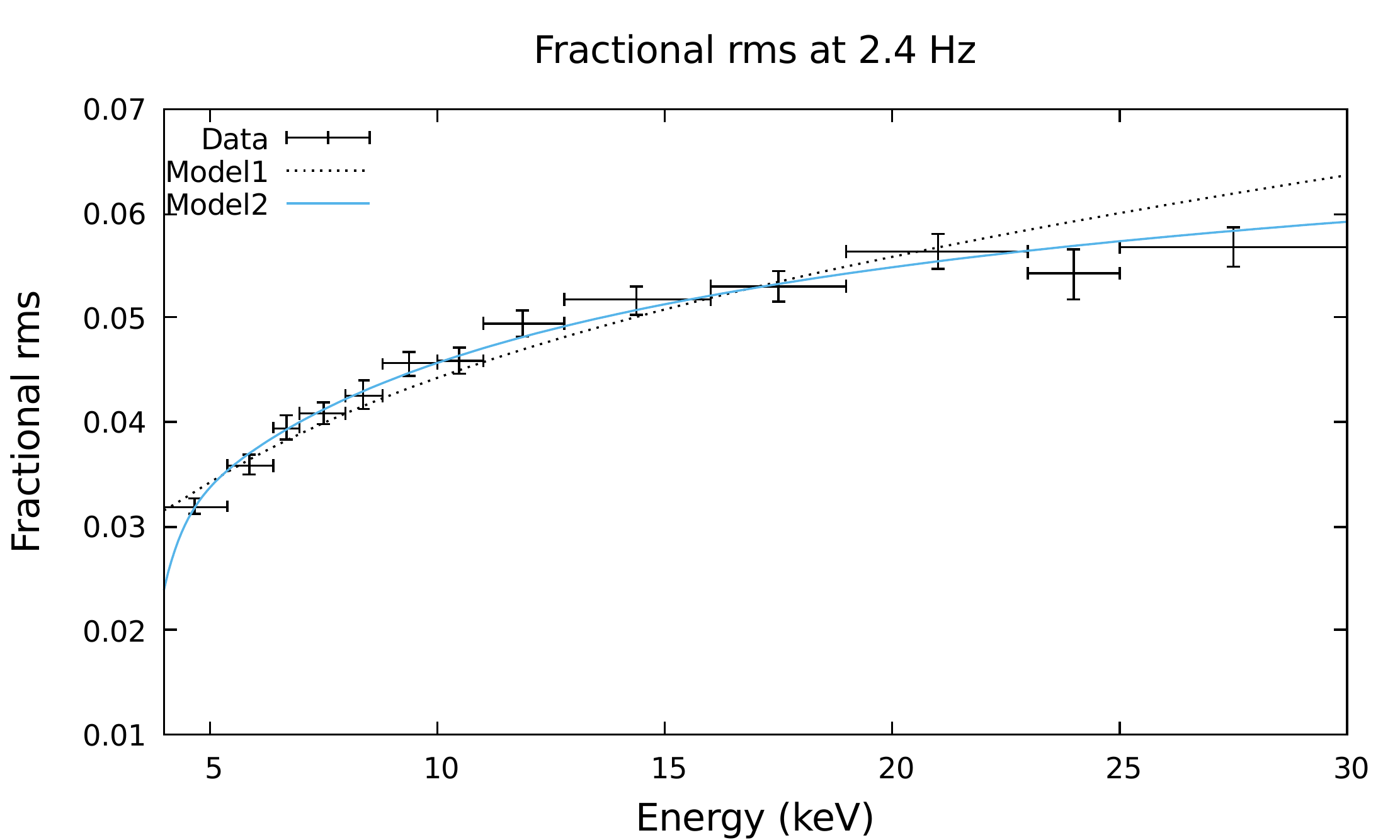}\hfill
\includegraphics[width=0.48\textwidth,height=4.8cm]{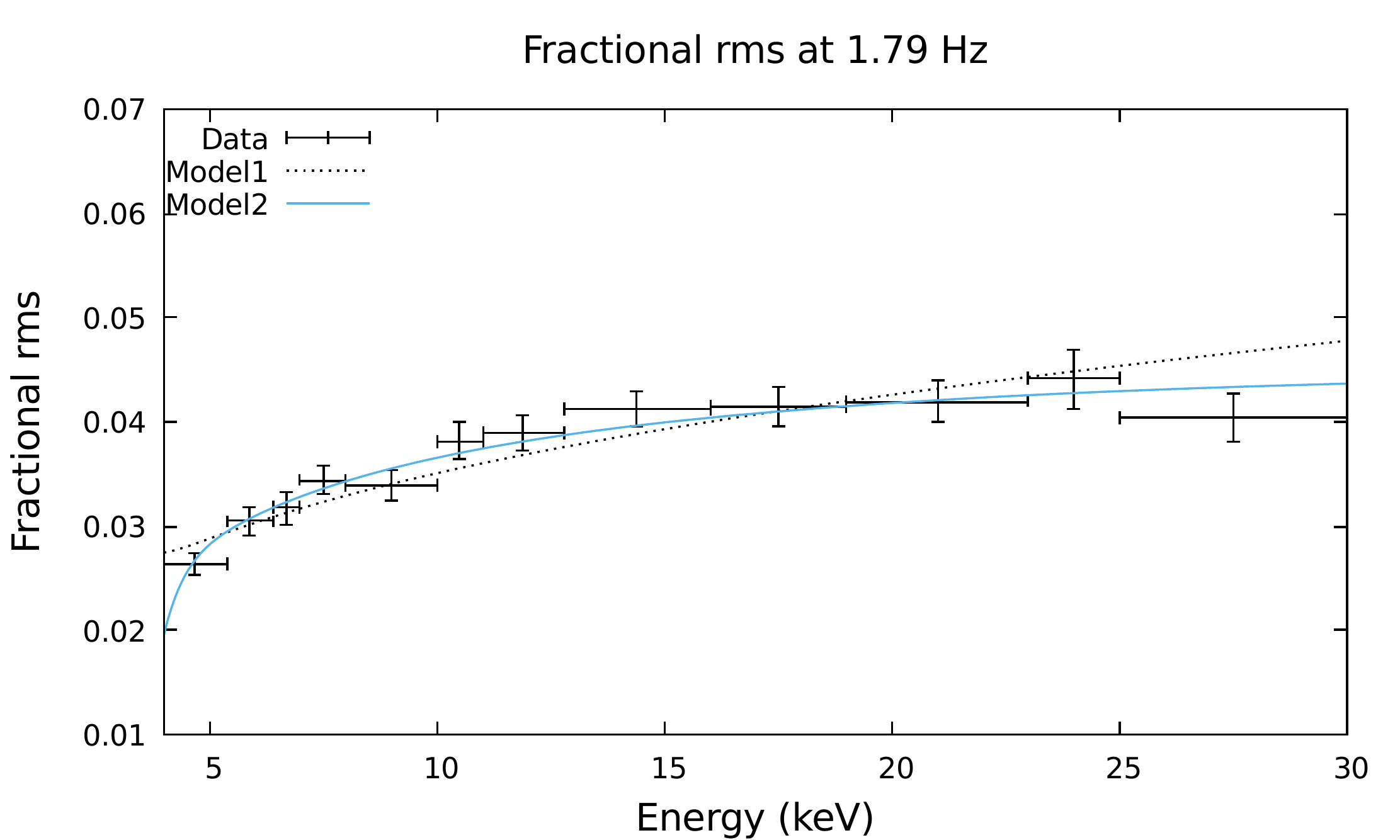}\hfill
\includegraphics[width=0.48\textwidth,height=4.8cm]{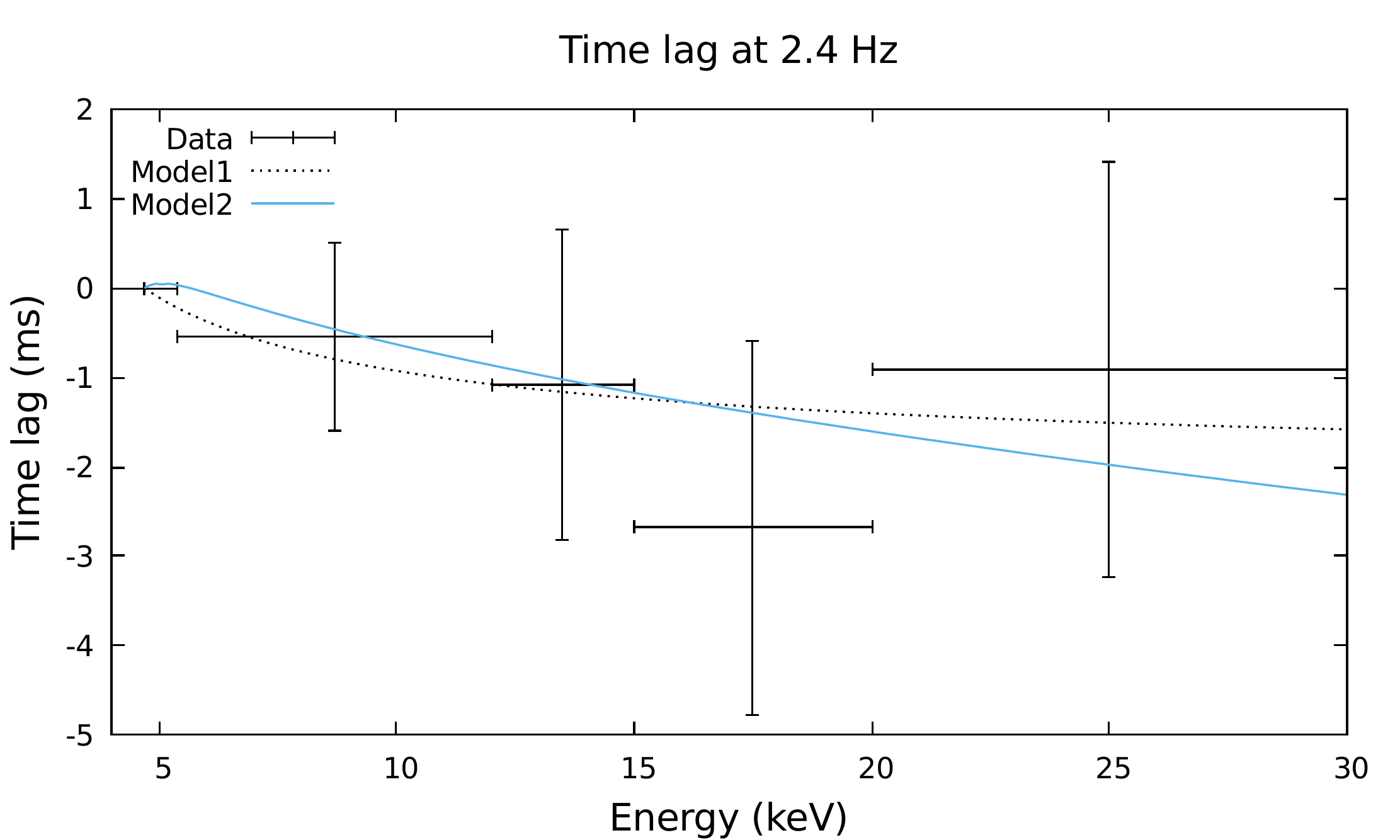}\hfill
\includegraphics[width=0.48\textwidth,height=4.8cm]{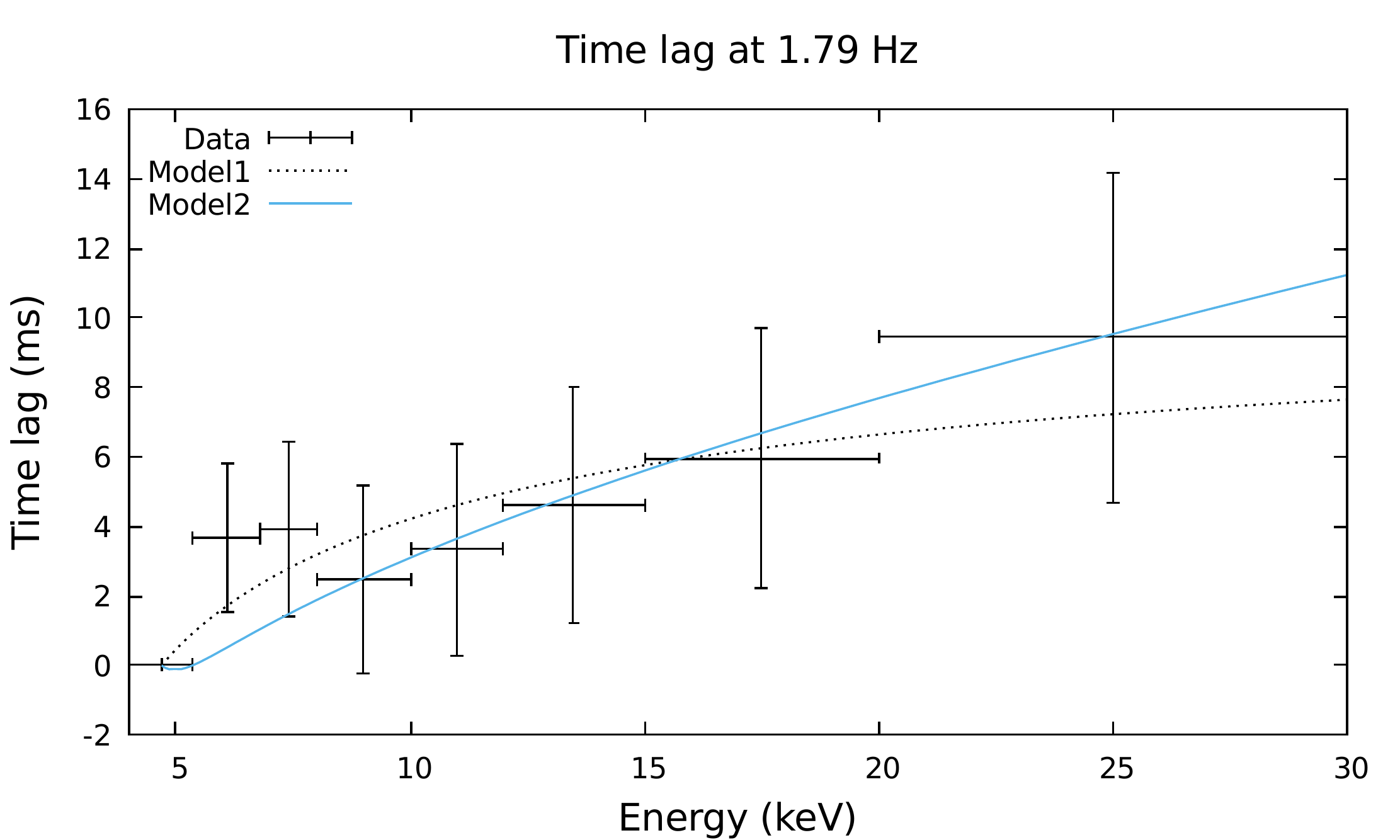}\hfill

\caption{Top and bottom panels show observed Fractional r.m.s and time-lags versus energy for two segments being fitted with Model 1 (dotted) and Model 2 (solid line) in the 4.0-30.0 keV respectively. Left panels correspond to segment with frequency greater than 2.2 Hz whereas right ones are for frequency less than 2.2 Hz.}
\label{fig:rl2}
\end{figure*}

Though the technique can be applied to any general spectral components,  \cite{2020MNRAS.498.2757G} considered a simple case of the X-ray emission spectrum arising out of a truncated accretion disc geometry, i.e. a combination of soft emission from geometrically thin and optically thick accretion disc truncated at a certain radius and hard emission from a coronal cloud of hot electrons in the vicinity of the compact object. In XSPEC, the former is represented by multicolor disc component - \textit{diskbb} and the latter by thermal comptonization - \textit{nthcomp}. \textit{diskbb} has two parameters - normalisation ($N_{disc}$) and inner disc temperature ($KT_{in}$) whereas \textit{nthcomp} has four parameters - normalisation ($N_c$), electron temperature ($KT_e$), spectral index ($\Gamma$) and seed photon temperature ($KT_{bb}$). It was further demonstrated in their work, how these spectral parameters can be transformed to a set of physically relevant ones - $N_{disc}$, $KT_{in}$, seed photon flux ($f_{sc}*N_{disc}$), heating rate ($\Dot{H}$) and optical depth ($\tau$). Note that the \textit{diskbb} parameters $N_{disc}$ and $KT_{in}$ are not transformed because their variations have a direct correspondence to changes in the inner disc radii and accretion rate respectively. 

In this present work for MAXI J1535-571, although we assume the same geometry of a truncated accretion disc, we now model the thermal comptonization component in energy spectrum using the convolution XSPEC model \textit{ThComp} instead of  \textit{nthcomp} as described in the previous section. The model \textit{ThComp} used the optical depth, scattering fraction and the electron temperature as the fitting parameters. This is convenient since now only the electron temperature $KT_e$ has to be translated to the heating rate of the corona, $\Dot{H}$  using the iterative method described in \cite{2020MNRAS.498.2757G} (see their equation(2)). Thus we have five possible physical parameters whose variation can lead to the observed energy dependent variability which are the inner disc temperature, $kT_{in}$, the normalization of the disc component, $N_{disc}$, the heating rate of the corona, $\Dot{H}$, the optical depth, $\tau$ and the scattering fraction $f_{sc}$. In the linear approximation, the variability of each of these parameter, $X$ over a mean value $<X_o>$ can be denoted by $\delta X = \Delta X/X_o$. The inner disc radius being  $R_{in} \propto N_{disc}^{1/2}$ and the accretion rate $\dot M \propto T_{in}^4 R_{in}^3$ implies that $\delta R_{in} = (1/2)\delta N_{disc}$ and  $\delta \dot M = 4 \delta kT_{in} + 3 \delta R_{in}$. Thus, instead of $\delta N_{disc}$ and $\delta kT_{in}$ it is convenient to use $\delta \dot M$ and $\delta R_{in}$ which along wth $\delta \dot H$, $\delta \tau$ and $\delta f_s$ specify the time dependent behaviour of the system.

\begin{figure*}

\centering
\includegraphics[width=0.48\textwidth,height=4.8cm]{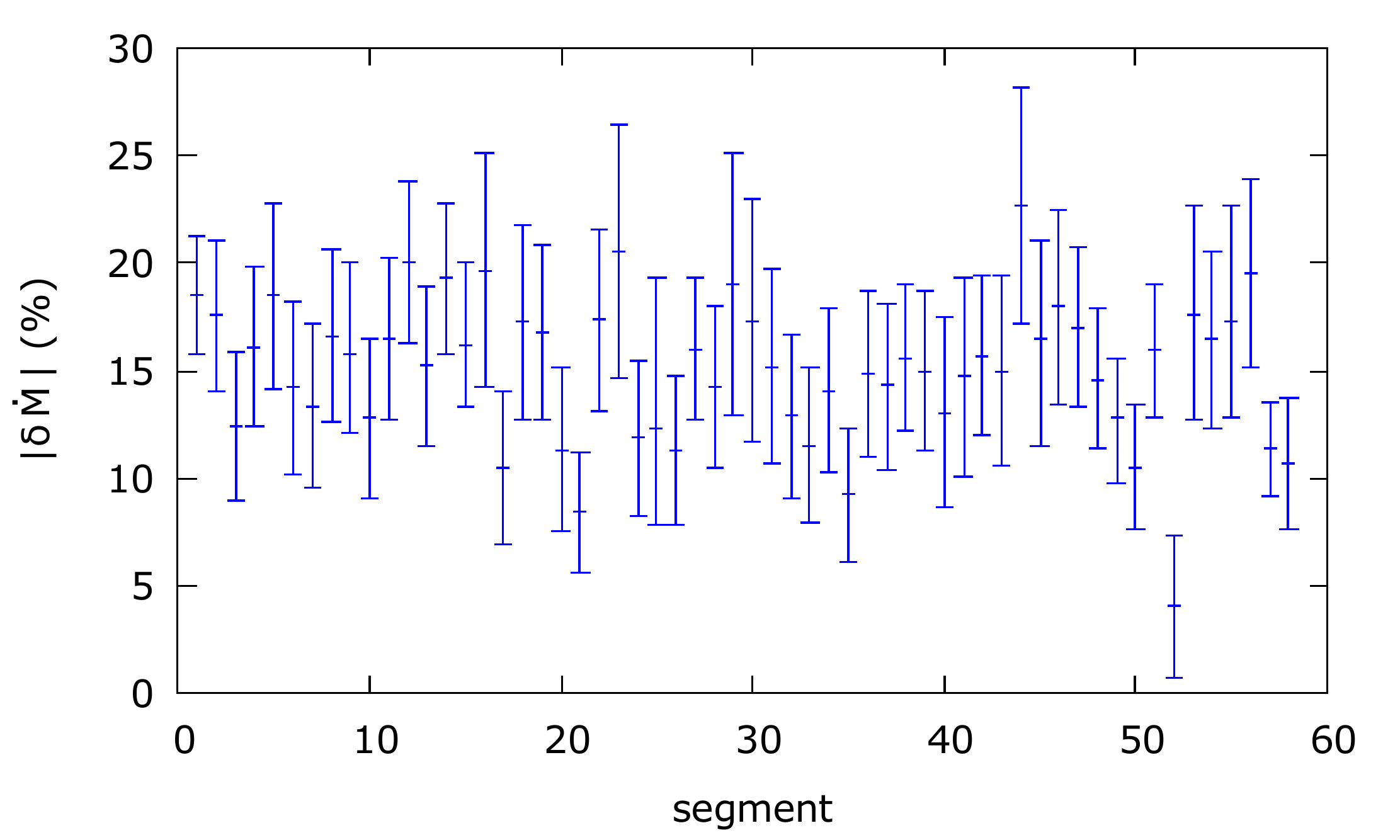}\hfill
\includegraphics[width=0.48\textwidth,height=4.8cm]{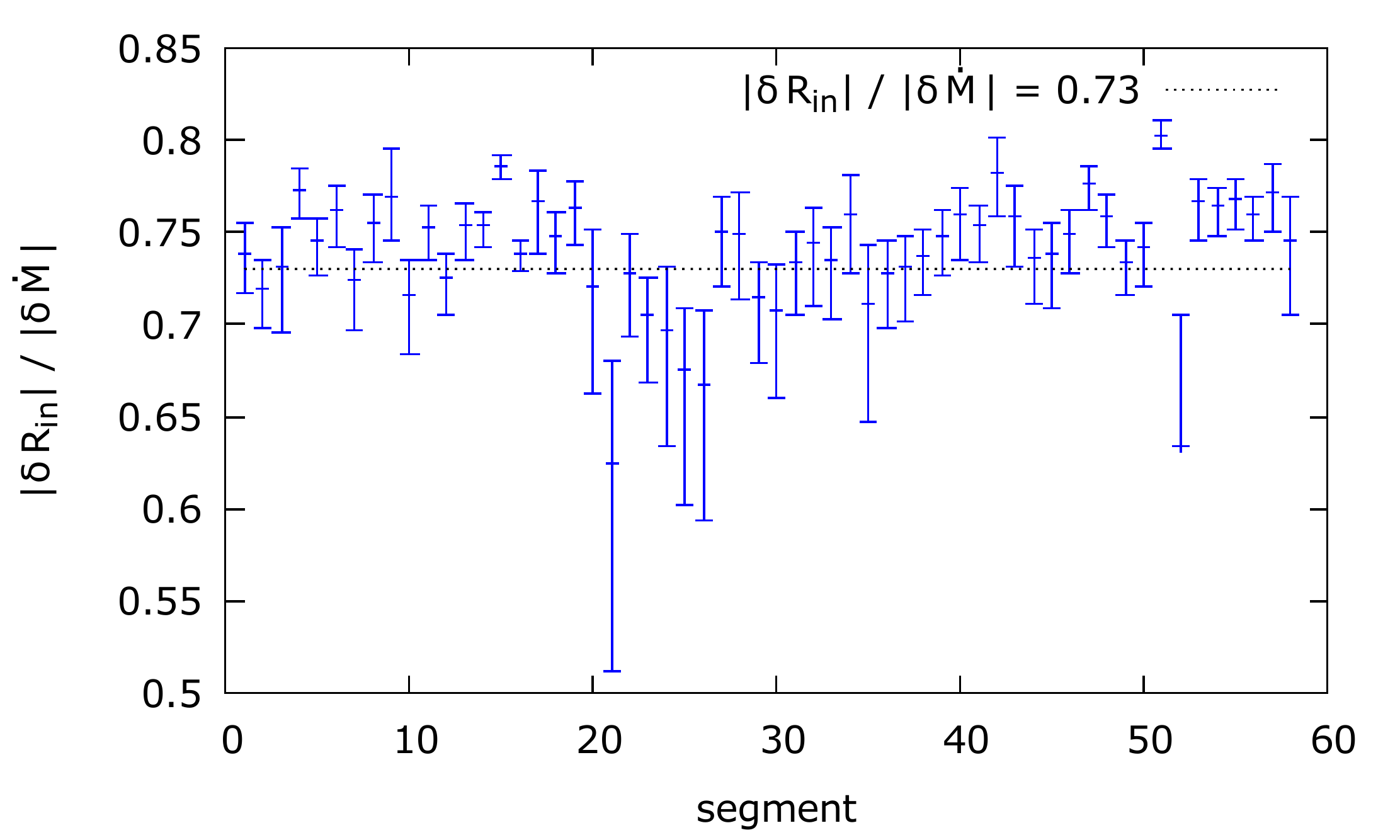}\hfill
\includegraphics[width=0.48\textwidth,height=4.8cm]{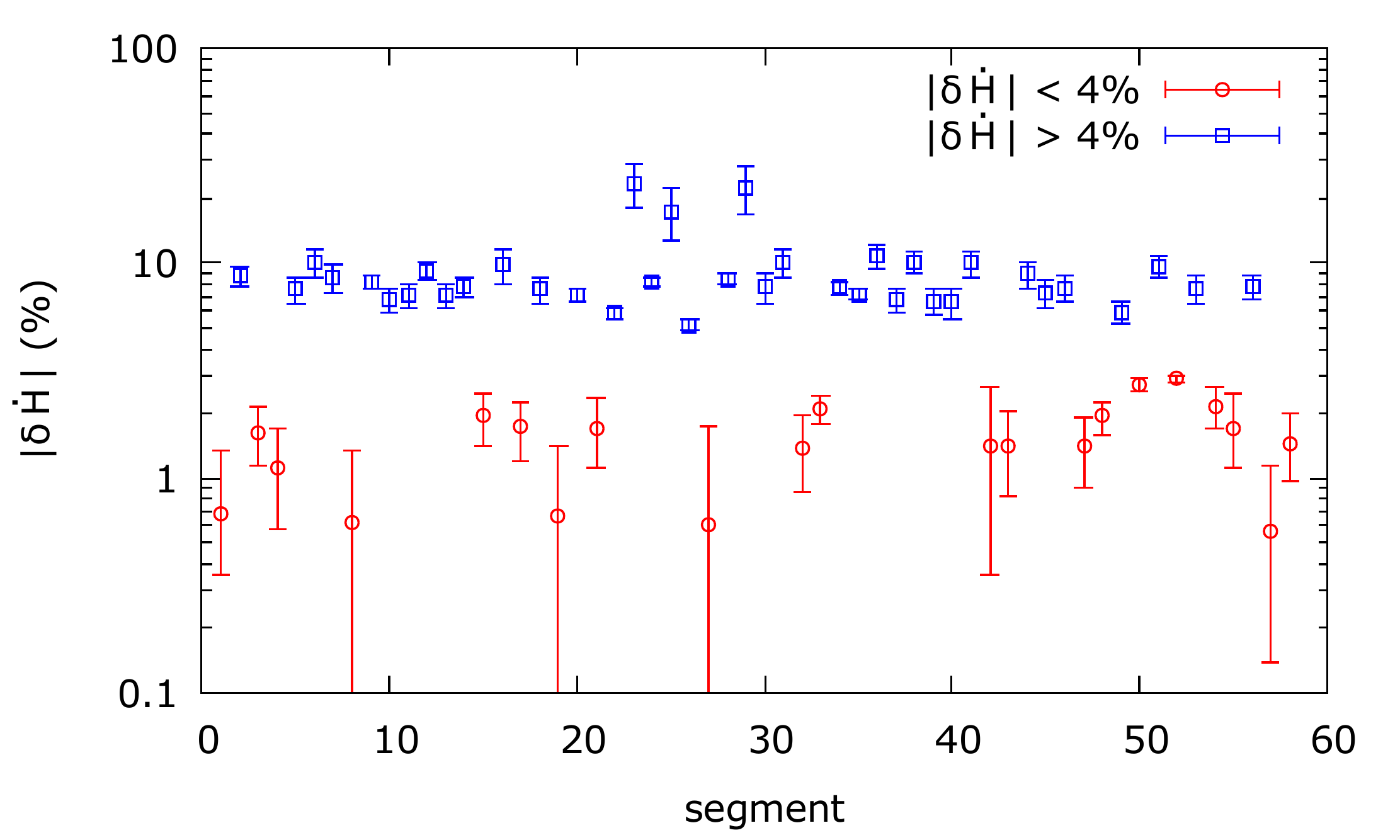}\hfill
\includegraphics[width=0.48\textwidth,height=4.8cm]{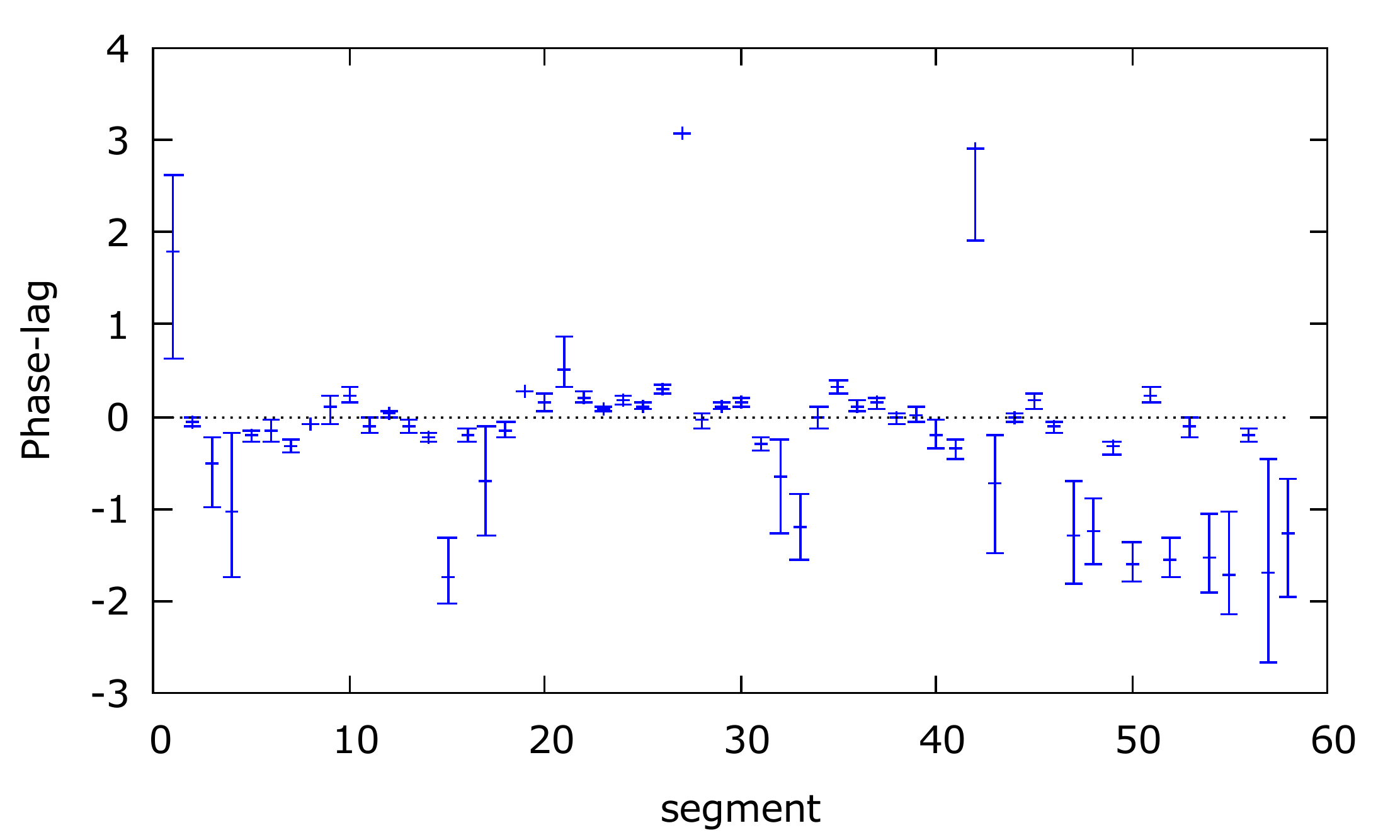}\hfill
\includegraphics[width=0.48\textwidth,height=4.8cm]{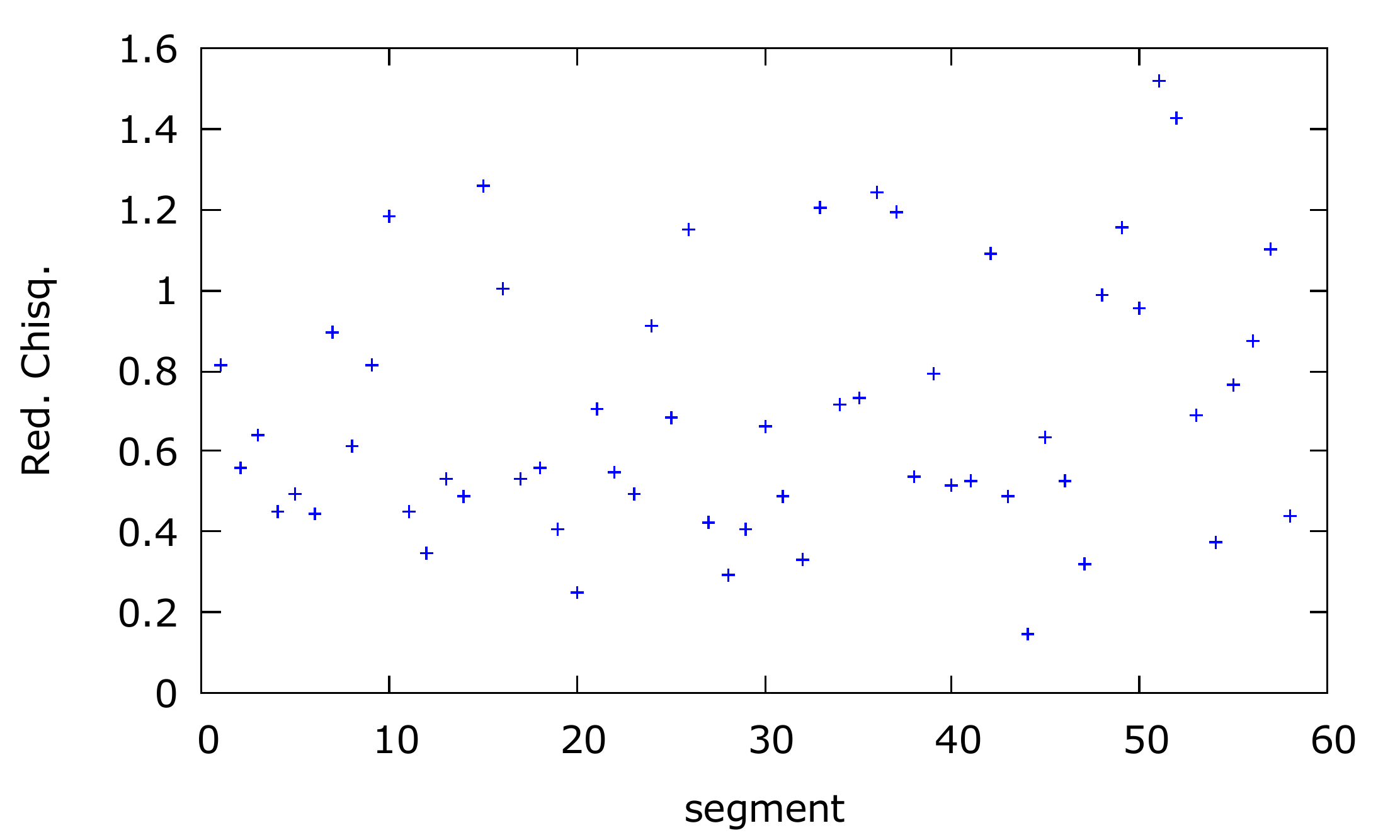}

\caption{Time evolution of Model~2 parameters. All the errors are at $1\sigma$ level.}
\label{fig:rlall2}
\end{figure*}

\begin{figure*}

\centering
\includegraphics[width=0.33\textwidth,height=4.8cm]{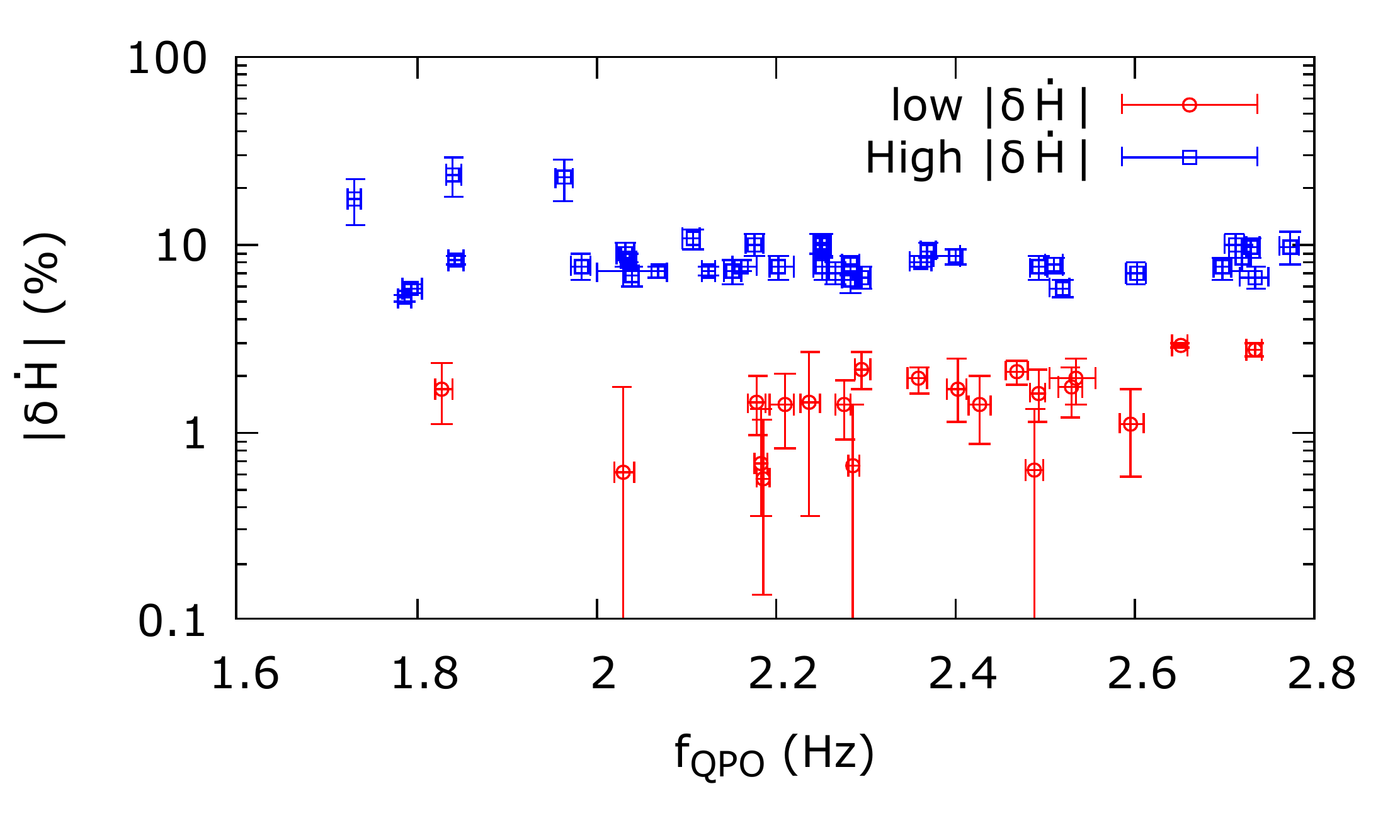}\hfill
\includegraphics[width=0.33\textwidth,height=4.8cm]{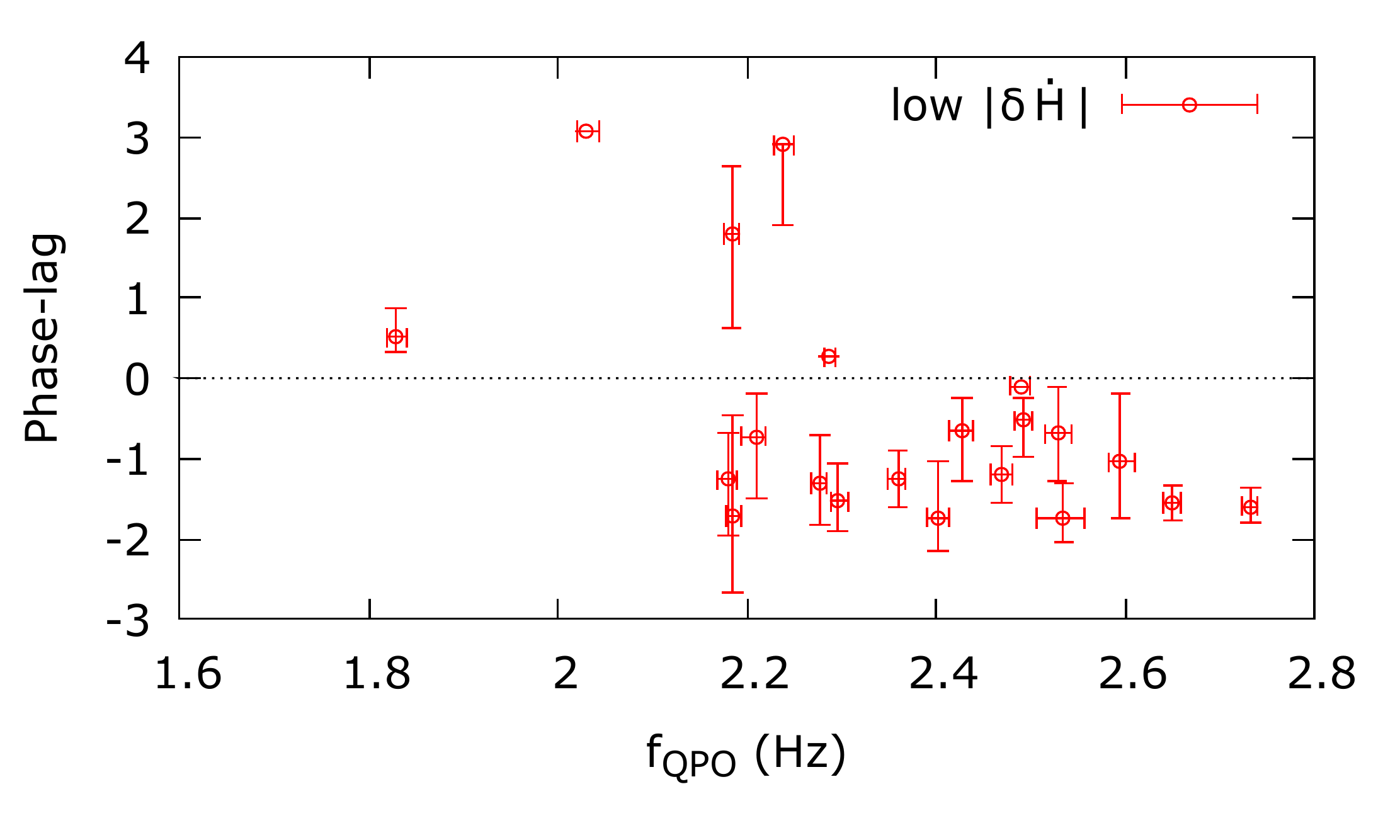}\hfill
\includegraphics[width=0.33\textwidth,height=4.8cm]{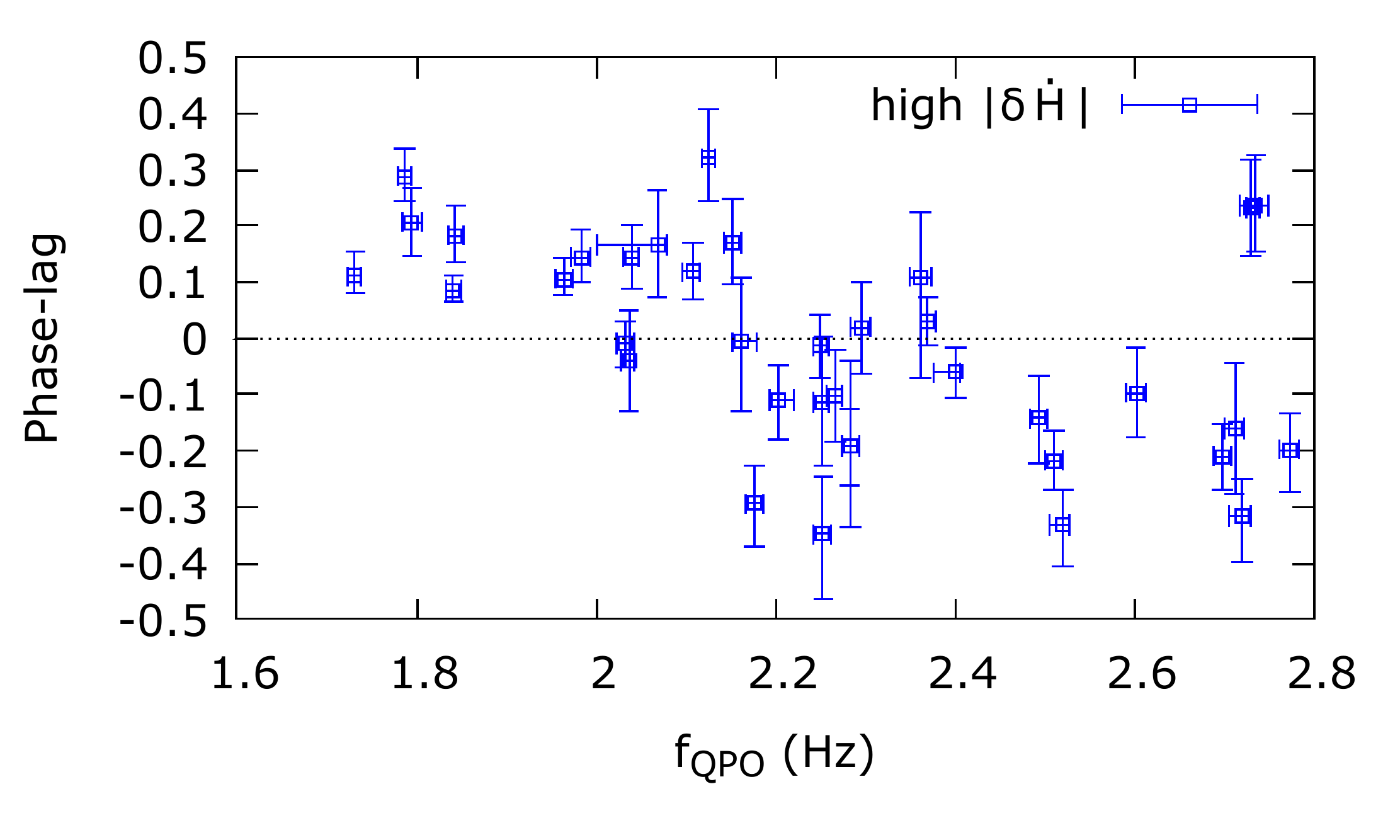}

\caption{Left panel shows correlation of $|\delta \Dot{H}|$ with QPO frequency. Middle and right panel show correlation of phase-lag between $\delta \Dot{H}$ and $\delta \dot M$ with QPO frequency for low $|\delta \dot H|$ and high $|\delta \dot H|$ respectively.}
\label{fig:f2pl}
\end{figure*}

Following \cite{2020MNRAS.498.2757G}, we started with the simple scenario (Model~1) where an accretion rate variation, ($\delta \dot M = |\delta \dot M| e^{i \omega t}$) at the QPO frequency ($\omega = 2\pi f_{QPO}$) induces a variation in the heating rate after a phase $\phi$, $\delta \Dot H = |\delta \Dot H| e^{i (\omega t-\phi)}$ with the rest of the parameters being constant. The three parameters $|\delta \dot M|$, $|\delta \Dot H|$ and $\phi$ determine the energy dependent fractional r.m.s and time-lag and the best fits for two representative segments are shown as dotted lines in Figure~\ref{fig:rl2}. The chi-square/(degrees of freedom) for the fitting are $27.2/14 \sim 1.9$ and $24.1/14 \sim 1.7$. The reduced chi-square for the segments ranged from 0.5 to 3.5 with typical values of 1.5.

Next, we considered the possibility that the inner truncated disc radius responds to the accretion rate variation with negligible delay i.e. $\delta R_{in} = |\delta R_{in}| e^{i \omega t}$.   With one extra parameter as the ratio of the inner radius and accretion rate variations, $ |\delta R_{in}|/|\delta \dot M|$, this scheme (Model 2) fits the data better than Model 1 by giving a reduced chi-square (/dof) of $7.24/13 (\sim 0.6)$ and $5.89/13 (\sim 0.5)$ for the two representative segments as shown by the solid lines in Figure~\ref{fig:rl2}. The reduced chi-square for the different segments ranged from 0.2 to 1.4 (bottom panel of Figure~\ref{fig:rlall2}) and hence we refrain from exploring more complex models in order not to over-model the system.  The variations of the four parameters ($|\delta \dot M|$, $|\delta R_{in}|/|\delta \dot M|$, $|\delta \Dot H|$, $\phi$) with segment number are shown in the panels of Figure~\ref{fig:rlall2}.

We start with noting that the ratio, $|\delta R_{in}| / |\delta \dot M|$ is nearly same and $ \sim 0.7$ for all segments which implies that $R_{in} \propto \dot M^{0.7}$. Thus, for the QPO we find that the inner disc radius positively correlates with the accretion rate. We note that any negative value of the ratio, which would imply an anti-correlation between the radius and accretion rate, yields a poor fit with large chi-square. The variation of the coronal heating rate $\dot H$ has a wide range of values ranging from $1.0$\% to as high as $30$\%. There seems to be a bi-modal distribution and thus we consider two samples with high ($> 4$\%) and low ($< 4$\%) values of $\delta \dot H$.  In Figure~\ref{fig:rlall2} (middle left panel) the two samples are plotted using red circles and blue squares respectively. The difference in the amplitude of $\delta \dot H$ is reflected in the middle and right panels of Figure~\ref{fig:f2pl} showing phase, $\phi$ as a function of QPO frequencies for the low and high $\delta \dot H$ samples respectively. High $\delta \dot H$ sample has smaller values of phase (implying shorter time-lags) while it is larger for low $\delta \dot H$ sample.

As shown in the left panel of Figure~\ref{fig:f2pl}, $\delta \dot H$ does not seem to depend on the QPO frequency while both samples consisting of high and low values of $\delta \dot H$ exhibit nearly the full range of QPO frequencies. For the sample with low values of $\delta \dot H$, the phase is not well constrained and has large error bars, but for the high values one can see an interesting result.  The phase is positive for frequencies less than $\sim 2.2$ Hz and negative for larger values. The bottom left panel of Figure~\ref{fig:rl2} show a representative fitting for QPO with frequency greater than 2.2 Hz (2.4 Hz), while the bottom right one is for the case when it is lower (1.79 Hz). The time-lag decreases as a function of energy (soft lag) for the higher frequencies while it increases (hard lag) for lower ones. Interestingly, this change in sign of the phase, i.e. switching from hard lags to soft ones  at $\sim$2-2.2 Hz was also found by \cite{2000ApJ...541..883R,2013MNRAS.436.2334P} and \cite{2020MNRAS.494.1375Z} for low-frequency QPOs in GRS 1915+105. Here, we find the same kind of behaviour for MAXI J1535-571 and a transition happening nearly at the same frequency. We further quantify this phenomenon as being a time-lag reversal between the variation of the accretion and the coronal heating.

\section{Discussion and Conclusions}

MAXI J1535-571, a black hole transient was detected in 2017 when it went into an outburst. Several satellite missions observed different phases of the outburst and revealed spectral and temporal characteristics similar to other  BHXBs. \textit{AstroSat} also looked at the source during the rising phase of outburst when it was in the hard-intermediate state and as expected for this state, prominent type-C QPOs were present in the power density spectra, whose frequency was found to be correlated with the high energy spectral index \citep{2019MNRAS.488..720B}. In this paper, we have studied the evolution of other spectral parameters such as the accretion rate and the inner radii of the disc and the scattering fraction. We found a  correlation between the QPO frequency and scattering fraction which is opposite to the anti-correlation reported by \cite{2013MNRAS.429.2655S} for other black hole binaries like GX 339-4, H1743-322 and XTE J1650-550 using RXTE data. \cite{2020ApJ...889L..36M}) and \cite{2021ApJ...909...63L} have shown that for the black hole system GRS 1915+105, the QPO frequency divided by the accretion rate has a functional dependence on the inner disc radius, allowing them to identify the frequency as the sound crossing time-scale in General Relativity. For an assumed distance, black hole mass, inclination angle and color factor in our analysis, we did find the QPO frequency divided by the accretion rate to be $\propto R_{in}^{-2.5}$. 
However it should be noted that this is an highly absorbed source in the hard state and the spectral band being covered is only from 1 keV onwards while the inner disc temperature is around 0.2 keV, so there may be doubts regarding the spectral fitting and specifically the disc parameters may not be accurate.

The primary motivation of this work, is to compute the energy dependent time-lag and r.m.s of the QPO at different freqencies and model them using a propagation model developed by \cite{2020MNRAS.498.2757G}. First, we discovered that the time-lag between the 12-15 keV photons with respect to 4-5.5 keV ones are positive when the QPO frequency is less than $\sim 2.2$ Hz and negative for larger frequencies. It is remarkable that the same phenomenon has been observed for a different black hole system GRS 1915+105 (\cite{2000ApJ...541..883R,2013MNRAS.436.2334P} and \cite{2020MNRAS.494.1375Z}). Although the two systems show different spectral characteristics during the QPO observations, the QPO frequency at which the time-lag flips sign is nearly the same.

Following, \cite{2020MNRAS.498.2757G}, we modelled the energy dependent time-lag and r.m.s using a general linear scheme where physical spectral parameters are varied coherently with the possibility of a time-lag between them. The simplest model which explains the data from all the segments is one where the accretion rate and the inner radius of the disc vary coherently with no time-lag and a coherent variation of the heating rate of the corona with a time lag. For the QPO, we can infer a dependence of the inner disc radius to the accretion rate to be $\dot M \propto R_{in}^{0.7}$, which is to be contrasted with the dependence found in the long term spectral evolution of the source where $\dot M \propto R_{in}^{2.4}$. This implies that the variation of the inner radii to the accretion rate is suppressed at the fast QPO time-scale compared to the long term behaviour. The positive correlation between inner disc radius and accretion rate has also been seen for the heart-beat oscillation of GRS 1915+105 (\cite{2022MNRAS.511.1841R,2012ApJ...750...71N}). On still longer time-scale of days corresponding to the evolution of the black hole transients, the inner disc radius decreases with accretion rate.

The time-lag between the variation of the coronal heating rate $\delta \dot H$ and the accretion rate $\delta \dot M$ changes sign when the QPO frequency crosses $2.2$ Hz, which is expected given the lag reversal seen between the hard and soft X-rays. The time-lags inferred are in the order of milli-seconds which is significantly smaller than the viscous or mass inflow time-scale. Hence, the time-delay between the disc and the corona maybe due to propagation of sound waves (\cite{2020ApJ...889L..36M}) or due to light travel times effects (\cite{2021MNRAS.503.5522K}). In any case, the direct and simplistic interpretation would be that for QPOs with frequency less than $2.2$ Hz, the dynamic origin is in the disc and then subsequently propagates inwards to the corona while for high frequencies the QPO's dynamic origin seems to be in the corona, which then propagates outwards to the disc. However, the scenario could be more complex with both the QPO originates in both the disc and corona and propagates to the other component after a time-delay. In that case, a net time-delay is observed which could be positive or negative depending on which is more dominant. Clearly more detailed models need to be developed to have a better understanding of the phenomenon, which will lead to a substantial improvement in our knowledge of these systems.

\section*{Acknowledgements}

We would like to thank the anonymous reviewer for constructive comments. This research work is utilizing data from \textbf{LAXPC} and \textbf{SXT} payloads on-board \textit{AstroSat}, available at Indian Space Science Data Centre (ISSDC). The work has made use of software provided by High Energy Astrophysics Science Archive Research Center (HEASARC). We are thankful to \textbf{LAXPC} Payload Operation Center (POC) and \textbf{SXT} POC at TIFR, Mumbai for providing software for data analysis. AG acknowledges the periodic visits to Inter-University Centre for Astronomy and Astrophysics (IUCAA), Pune to carry out the major part of the work. AG is thankful to Dr. Yash Bhargava, IUCAA for useful discussion regarding segmentation of data and spectral analysis. AG thanks UGC, Govt. of India for providing fellowship under the UGC-JRF scheme (Ref. No.: 1449/CSIR NET JUNE 2019). We also acknowledge the financial aid by ISRO under AstroSat data utilization AO (No. DS$\_$2B-13013(2)/2/2022-sec.2 dated Feb 17, 2022).

\section*{Data Availability}

 The data utilized in this article are available at \textit{Astrosat}-ISSDC website (\url{http://astrobrowse.issdc.gov.in/astro\_archive/archive}). The Fortran code underlying the article can be accessed by a reasonable request to the corresponding author.



\bibliographystyle{mnras}
\bibliography{Project2} 

\begin{thebibliography}{}
\makeatletter
\relax
\def\mn@urlcharsother{\let\do\@makeother \do\$\do\&\do\#\do\^\do\_\do\%\do\~}
\def\mn@doi{\begingroup\mn@urlcharsother \@ifnextchar [ {\mn@doi@}
  {\mn@doi@[]}}
\def\mn@doi@[#1]#2{\def\@tempa{#1}\ifx\@tempa\@empty \href
  {http://dx.doi.org/#2} {doi:#2}\else \href {http://dx.doi.org/#2} {#1}\fi
  \endgroup}
\def\mn@eprint#1#2{\mn@eprint@#1:#2::\@nil}
\def\mn@eprint@arXiv#1{\href {http://arxiv.org/abs/#1} {{\tt arXiv:#1}}}
\def\mn@eprint@dblp#1{\href {http://dblp.uni-trier.de/rec/bibtex/#1.xml}
  {dblp:#1}}
\def\mn@eprint@#1:#2:#3:#4\@nil{\def\@tempa {#1}\def\@tempb {#2}\def\@tempc
  {#3}\ifx \@tempc \@empty \let \@tempc \@tempb \let \@tempb \@tempa \fi \ifx
  \@tempb \@empty \def\@tempb {arXiv}\fi \@ifundefined
  {mn@eprint@\@tempb}{\@tempb:\@tempc}{\expandafter \expandafter \csname
  mn@eprint@\@tempb\endcsname \expandafter{\@tempc}}}

\bibitem[\protect\citeauthoryear{{Agrawal} et~al.,}{{Agrawal}
  et~al.}{2017}]{2017JApA...38...30A}
{Agrawal} P.~C.,  et~al., 2017, \mn@doi [Journal of Astrophysics and Astronomy]
  {10.1007/s12036-017-9451-z}, \href
  {https://ui.adsabs.harvard.edu/abs/2017JApA...38...30A} {38, 30}

\bibitem[\protect\citeauthoryear{{Belloni} \& {Motta}}{{Belloni} \&
  {Motta}}{2016}]{2016ASSL..440...61B}
{Belloni} T.~M.,  {Motta} S.~E.,  2016, {Transient Black Hole Binaries}.
p.~61, \mn@doi{10.1007/978-3-662-52859-4\_2}

\bibitem[\protect\citeauthoryear{{Belloni}, {Motta}  \&
  {Mu{\~n}oz-Darias}}{{Belloni} et~al.}{2011}]{2011BASI...39..409B}
{Belloni} T.~M.,  {Motta} S.~E.,   {Mu{\~n}oz-Darias} T.,  2011, Bulletin of
  the Astronomical Society of India, \href
  {https://ui.adsabs.harvard.edu/abs/2011BASI...39..409B} {39, 409}

\bibitem[\protect\citeauthoryear{{Bhargava}, {Belloni}, {Bhattacharya}  \&
  {Misra}}{{Bhargava} et~al.}{2019}]{2019MNRAS.488..720B}
{Bhargava} Y.,  {Belloni} T.,  {Bhattacharya} D.,   {Misra} R.,  2019, \mn@doi
  [\mnras] {10.1093/mnras/stz1774}, \href
  {https://ui.adsabs.harvard.edu/abs/2019MNRAS.488..720B} {488, 720}

\bibitem[\protect\citeauthoryear{{Chakrabarti} \& {Manickam}}{{Chakrabarti} \&
  {Manickam}}{2000}]{2000ApJ...531L..41C}
{Chakrabarti} S.~K.,  {Manickam} S.~G.,  2000, \mn@doi [\apjl]
  {10.1086/312512}, \href
  {https://ui.adsabs.harvard.edu/abs/2000ApJ...531L..41C} {531, L41}

\bibitem[\protect\citeauthoryear{{Frank}, {King}  \& {Raine}}{{Frank}
  et~al.}{2002}]{2002apa}
{Frank} J.,  {King} A.,   {Raine} D.~J.,  2002, {Accretion Power in
  Astrophysics: Third Edition}

\bibitem[\protect\citeauthoryear{{Garg}, {Misra}  \& {Sen}}{{Garg}
  et~al.}{2020}]{2020MNRAS.498.2757G}
{Garg} A.,  {Misra} R.,   {Sen} S.,  2020, \mn@doi [\mnras]
  {10.1093/mnras/staa2506}, \href
  {https://ui.adsabs.harvard.edu/abs/2020MNRAS.498.2757G} {498, 2757}

\bibitem[\protect\citeauthoryear{{Ingram} \& {Motta}}{{Ingram} \&
  {Motta}}{2019}]{2019NewAR..8501524I}
{Ingram} A.~R.,  {Motta} S.~E.,  2019, \mn@doi [\nar]
  {10.1016/j.newar.2020.101524}, \href
  {https://ui.adsabs.harvard.edu/abs/2019NewAR..8501524I} {85, 101524}

\bibitem[\protect\citeauthoryear{{Ingram}, {Done}  \& {Fragile}}{{Ingram}
  et~al.}{2009}]{2009MNRAS.397L.101I}
{Ingram} A.,  {Done} C.,   {Fragile} P.~C.,  2009, \mn@doi [\mnras]
  {10.1111/j.1745-3933.2009.00693.x}, \href
  {https://ui.adsabs.harvard.edu/abs/2009MNRAS.397L.101I} {397, L101}

\bibitem[\protect\citeauthoryear{{Jithesh}, {Maqbool}, {Misra}, {T}, {Mall}  \&
  {James}}{{Jithesh} et~al.}{2019}]{2019ApJ...887..101J}
{Jithesh} V.,  {Maqbool} B.,  {Misra} R.,  {T} A.~R.,  {Mall} G.,   {James} M.,
   2019, \mn@doi [\apj] {10.3847/1538-4357/ab4f6a}, \href
  {https://ui.adsabs.harvard.edu/abs/2019ApJ...887..101J} {887, 101}

\bibitem[\protect\citeauthoryear{{Karpouzas}, {M{\'e}ndez}, {Garc{\'\i}a},
  {Zhang}, {Altamirano}, {Belloni}  \& {Zhang}}{{Karpouzas}
  et~al.}{2021}]{2021MNRAS.503.5522K}
{Karpouzas} K.,  {M{\'e}ndez} M.,  {Garc{\'\i}a} F.,  {Zhang} L.,  {Altamirano}
  D.,  {Belloni} T.,   {Zhang} Y.,  2021, \mn@doi [\mnras]
  {10.1093/mnras/stab827}, \href
  {https://ui.adsabs.harvard.edu/abs/2021MNRAS.503.5522K} {503, 5522}

\bibitem[\protect\citeauthoryear{{Liu}, {Ji}, {Bambi}, {Jain}, {Misra},
  {Rawat}, {Yadav}  \& {Zhang}}{{Liu} et~al.}{2021}]{2021ApJ...909...63L}
{Liu} H.,  {Ji} L.,  {Bambi} C.,  {Jain} P.,  {Misra} R.,  {Rawat} D.,  {Yadav}
  J.~S.,   {Zhang} Y.,  2021, \mn@doi [\apj] {10.3847/1538-4357/abdf65}, \href
  {https://ui.adsabs.harvard.edu/abs/2021ApJ...909...63L} {909, 63}

\bibitem[\protect\citeauthoryear{{Lyubarskii}}{{Lyubarskii}}{1997}]{1997MNRAS.292..679L}
{Lyubarskii} Y.~E.,  1997, \mn@doi [\mnras] {10.1093/mnras/292.3.679}, \href
  {https://ui.adsabs.harvard.edu/abs/1997MNRAS.292..679L} {292, 679}

\bibitem[\protect\citeauthoryear{{Maqbool} et~al.,}{{Maqbool}
  et~al.}{2019}]{2019MNRAS.486.2964M}
{Maqbool} B.,  et~al., 2019, \mn@doi [\mnras] {10.1093/mnras/stz930}, \href
  {https://ui.adsabs.harvard.edu/abs/2019MNRAS.486.2964M} {486, 2964}

\bibitem[\protect\citeauthoryear{{Mir}, {Misra}, {Pahari}, {Iqbal}  \&
  {Ahmad}}{{Mir} et~al.}{2016}]{2016MNRAS.457.2999M}
{Mir} M.~H.,  {Misra} R.,  {Pahari} M.,  {Iqbal} N.,   {Ahmad} N.,  2016,
  \mn@doi [\mnras] {10.1093/mnras/stw156}, \href
  {https://ui.adsabs.harvard.edu/abs/2016MNRAS.457.2999M} {457, 2999}

\bibitem[\protect\citeauthoryear{{Misra} \& {Mandal}}{{Misra} \&
  {Mandal}}{2013}]{2013ApJ...779...71M}
{Misra} R.,  {Mandal} S.,  2013, \mn@doi [\apj] {10.1088/0004-637X/779/1/71},
  \href {https://ui.adsabs.harvard.edu/abs/2013ApJ...779...71M} {779, 71}

\bibitem[\protect\citeauthoryear{{Misra}, {Rawat}, {Yadav}  \& {Jain}}{{Misra}
  et~al.}{2020}]{2020ApJ...889L..36M}
{Misra} R.,  {Rawat} D.,  {Yadav} J.~S.,   {Jain} P.,  2020, \mn@doi [\apjl]
  {10.3847/2041-8213/ab6ddc}, \href
  {https://ui.adsabs.harvard.edu/abs/2020ApJ...889L..36M} {889, L36}

\bibitem[\protect\citeauthoryear{{Mitsuda} et~al.,}{{Mitsuda}
  et~al.}{1984}]{1984PASJ...36..741M}
{Mitsuda} K.,  et~al., 1984, \pasj, \href
  {https://ui.adsabs.harvard.edu/abs/1984PASJ...36..741M} {36, 741}

\bibitem[\protect\citeauthoryear{{Molteni}, {Sponholz}  \&
  {Chakrabarti}}{{Molteni} et~al.}{1996}]{1996ApJ...457..805M}
{Molteni} D.,  {Sponholz} H.,   {Chakrabarti} S.~K.,  1996, \mn@doi [\apj]
  {10.1086/176775}, \href
  {https://ui.adsabs.harvard.edu/abs/1996ApJ...457..805M} {457, 805}

\bibitem[\protect\citeauthoryear{{Motta}}{{Motta}}{2016}]{2016AN....337..398M}
{Motta} S.~E.,  2016, \mn@doi [Astronomische Nachrichten]
  {10.1002/asna.201612320}, \href
  {https://ui.adsabs.harvard.edu/abs/2016AN....337..398M} {337, 398}

\bibitem[\protect\citeauthoryear{{Motta}, {Franchini}, {Lodato}  \&
  {Mastroserio}}{{Motta} et~al.}{2018}]{2018MNRAS.473..431M}
{Motta} S.~E.,  {Franchini} A.,  {Lodato} G.,   {Mastroserio} G.,  2018,
  \mn@doi [\mnras] {10.1093/mnras/stx2358}, \href
  {https://ui.adsabs.harvard.edu/abs/2018MNRAS.473..431M} {473, 431}

\bibitem[\protect\citeauthoryear{{Mudambi}, {Maqbool}, {Misra}, {Hebbar},
  {Yadav}, {Gudennavar}  \& {S.~G.}}{{Mudambi}
  et~al.}{2020}]{2020ApJ...889L..17M}
{Mudambi} S.~P.,  {Maqbool} B.,  {Misra} R.,  {Hebbar} S.,  {Yadav} J.~S.,
  {Gudennavar} S.~B.,   {S.~G.} B.,  2020, \mn@doi [\apjl]
  {10.3847/2041-8213/ab66bc}, \href
  {https://ui.adsabs.harvard.edu/abs/2020ApJ...889L..17M} {889, L17}

\bibitem[\protect\citeauthoryear{{Muno}, {Remillard}, {Morgan}, {Waltman},
  {Dhawan}, {Hjellming}  \& {Pooley}}{{Muno}
  et~al.}{2001}]{2001ApJ...556..515M}
{Muno} M.~P.,  {Remillard} R.~A.,  {Morgan} E.~H.,  {Waltman} E.~B.,  {Dhawan}
  V.,  {Hjellming} R.~M.,   {Pooley} G.,  2001, \mn@doi [\apj]
  {10.1086/321604}, \href
  {https://ui.adsabs.harvard.edu/abs/2001ApJ...556..515M} {556, 515}

\bibitem[\protect\citeauthoryear{{Neilsen}, {Remillard}  \& {Lee}}{{Neilsen}
  et~al.}{2012}]{2012ApJ...750...71N}
{Neilsen} J.,  {Remillard} R.~A.,   {Lee} J.~C.,  2012, \mn@doi [\apj]
  {10.1088/0004-637X/750/1/71}, \href
  {https://ui.adsabs.harvard.edu/abs/2012ApJ...750...71N} {750, 71}

\bibitem[\protect\citeauthoryear{{Nowak}, {Vaughan}, {Wilms}, {Dove}  \&
  {Begelman}}{{Nowak} et~al.}{1999}]{1999ApJ...510..874N}
{Nowak} M.~A.,  {Vaughan} B.~A.,  {Wilms} J.,  {Dove} J.~B.,   {Begelman}
  M.~C.,  1999, \mn@doi [\apj] {10.1086/306610}, \href
  {https://ui.adsabs.harvard.edu/abs/1999ApJ...510..874N} {510, 874}

\bibitem[\protect\citeauthoryear{{Pahari}, {Misra}, {Mukherjee}, {Yadav}  \&
  {Pandey}}{{Pahari} et~al.}{2013}]{2013MNRAS.436.2334P}
{Pahari} M.,  {Misra} R.,  {Mukherjee} A.,  {Yadav} J.~S.,   {Pandey} S.~K.,
  2013, \mn@doi [\mnras] {10.1093/mnras/stt1732}, \href
  {https://ui.adsabs.harvard.edu/abs/2013MNRAS.436.2334P} {436, 2334}

\bibitem[\protect\citeauthoryear{{Rana} \& {Mangalam}}{{Rana} \&
  {Mangalam}}{2020}]{2020ApJ...903..121R}
{Rana} P.,  {Mangalam} A.,  2020, \mn@doi [\apj] {10.3847/1538-4357/abb707},
  \href {https://ui.adsabs.harvard.edu/abs/2020ApJ...903..121R} {903, 121}

\bibitem[\protect\citeauthoryear{{Rawat}, {Misra}, {Jain}  \& {Yadav}}{{Rawat}
  et~al.}{2022}]{2022MNRAS.511.1841R}
{Rawat} D.,  {Misra} R.,  {Jain} P.,   {Yadav} J.~S.,  2022, \mn@doi [\mnras]
  {10.1093/mnras/stac154}, \href
  {https://ui.adsabs.harvard.edu/abs/2022MNRAS.511.1841R} {511, 1841}

\bibitem[\protect\citeauthoryear{{Reig}, {Belloni}, {van der Klis},
  {M{\'e}ndez}, {Kylafis}  \& {Ford}}{{Reig}
  et~al.}{2000}]{2000ApJ...541..883R}
{Reig} P.,  {Belloni} T.,  {van der Klis} M.,  {M{\'e}ndez} M.,  {Kylafis}
  N.~D.,   {Ford} E.~C.,  2000, \mn@doi [\apj] {10.1086/309469}, \href
  {https://ui.adsabs.harvard.edu/abs/2000ApJ...541..883R} {541, 883}

\bibitem[\protect\citeauthoryear{{Shakura} \& {Sunyaev}}{{Shakura} \&
  {Sunyaev}}{1973}]{1973A&A....24..337S}
{Shakura} N.~I.,  {Sunyaev} R.~A.,  1973, \aap, \href
  {https://ui.adsabs.harvard.edu/abs/1973A&A....24..337S} {500, 33}

\bibitem[\protect\citeauthoryear{{Singh} et~al.,}{{Singh}
  et~al.}{2017}]{2017JApA...38...29S}
{Singh} K.~P.,  et~al., 2017, \mn@doi [Journal of Astrophysics and Astronomy]
  {10.1007/s12036-017-9448-7}, \href
  {https://ui.adsabs.harvard.edu/abs/2017JApA...38...29S} {38, 29}

\bibitem[\protect\citeauthoryear{{Sreehari}, {Ravishankar}, {Iyer}, {Agrawal},
  {Katoch}, {Mandal}  \& {Nandi}}{{Sreehari}
  et~al.}{2019}]{2019MNRAS.487..928S}
{Sreehari} H.,  {Ravishankar} B.~T.,  {Iyer} N.,  {Agrawal} V.~K.,  {Katoch}
  T.~B.,  {Mandal} S.,   {Nandi} A.,  2019, \mn@doi [\mnras]
  {10.1093/mnras/stz1327}, \href
  {https://ui.adsabs.harvard.edu/abs/2019MNRAS.487..928S} {487, 928}

\bibitem[\protect\citeauthoryear{{Sridhar}, {Bhattacharyya}, {Chandra}  \&
  {Antia}}{{Sridhar} et~al.}{2019}]{2019MNRAS.487.4221S}
{Sridhar} N.,  {Bhattacharyya} S.,  {Chandra} S.,   {Antia} H.~M.,  2019,
  \mn@doi [\mnras] {10.1093/mnras/stz1476}, \href
  {https://ui.adsabs.harvard.edu/abs/2019MNRAS.487.4221S} {487, 4221}

\bibitem[\protect\citeauthoryear{{Stiele}, {Belloni}, {Kalemci}  \&
  {Motta}}{{Stiele} et~al.}{2013}]{2013MNRAS.429.2655S}
{Stiele} H.,  {Belloni} T.~M.,  {Kalemci} E.,   {Motta} S.,  2013, \mn@doi
  [\mnras] {10.1093/mnras/sts548}, \href
  {https://ui.adsabs.harvard.edu/abs/2013MNRAS.429.2655S} {429, 2655}

\bibitem[\protect\citeauthoryear{{Tagger} \& {Pellat}}{{Tagger} \&
  {Pellat}}{1999}]{1999A&A...349.1003T}
{Tagger} M.,  {Pellat} R.,  1999, \aap, \href
  {https://ui.adsabs.harvard.edu/abs/1999A&A...349.1003T} {349, 1003}

\bibitem[\protect\citeauthoryear{{Titarchuk} \& {Fiorito}}{{Titarchuk} \&
  {Fiorito}}{2004}]{2004ApJ...612..988T}
{Titarchuk} L.,  {Fiorito} R.,  2004, \mn@doi [\apj] {10.1086/422573}, \href
  {https://ui.adsabs.harvard.edu/abs/2004ApJ...612..988T} {612, 988}

\bibitem[\protect\citeauthoryear{{Van der Klis}}{{Van der
  Klis}}{1989}]{1989ASIC..262...27V}
{Van der Klis} M.,  1989, in {{\"O}gelman} H.,  {van den Heuvel} E.~P.~J.,
  eds,  NATO Advanced Study Institute (ASI) Series C Vol. 262, Timing Neutron
  Stars. p.~27

\bibitem[\protect\citeauthoryear{{Vignarca}, {Migliari}, {Belloni}, {Psaltis}
  \& {van der Klis}}{{Vignarca} et~al.}{2003}]{2003A&A...397..729V}
{Vignarca} F.,  {Migliari} S.,  {Belloni} T.,  {Psaltis} D.,   {van der Klis}
  M.,  2003, \mn@doi [\aap] {10.1051/0004-6361:20021542}, \href
  {https://ui.adsabs.harvard.edu/abs/2003A&A...397..729V} {397, 729}

\bibitem[\protect\citeauthoryear{{Wilms}, {Allen}  \& {McCray}}{{Wilms}
  et~al.}{2000}]{2000ApJ...542..914W}
{Wilms} J.,  {Allen} A.,   {McCray} R.,  2000, \mn@doi [\apj] {10.1086/317016},
  \href {https://ui.adsabs.harvard.edu/abs/2000ApJ...542..914W} {542, 914}

\bibitem[\protect\citeauthoryear{{Yadav} et~al.,}{{Yadav}
  et~al.}{2016}]{2016SPIE.9905E..1DY}
{Yadav} J.~S.,  et~al., 2016, in {den Herder} J.-W.~A.,  {Takahashi} T.,
  {Bautz} M.,  eds,  Society of Photo-Optical Instrumentation Engineers (SPIE)
  Conference Series Vol. 9905, Space Telescopes and Instrumentation 2016:
  Ultraviolet to Gamma Ray. p. 99051D, \mn@doi{10.1117/12.2231857}

\bibitem[\protect\citeauthoryear{{Zdziarski}, {Johnson}  \&
  {Magdziarz}}{{Zdziarski} et~al.}{1996}]{1996MNRAS.283..193Z}
{Zdziarski} A.~A.,  {Johnson} W.~N.,   {Magdziarz} P.,  1996, \mn@doi [\mnras]
  {10.1093/mnras/283.1.193}, \href
  {https://ui.adsabs.harvard.edu/abs/1996MNRAS.283..193Z} {283, 193}

\bibitem[\protect\citeauthoryear{{Zdziarski}, {Szanecki}, {Poutanen},
  {Gierli{\'n}ski}  \& {Biernacki}}{{Zdziarski}
  et~al.}{2020}]{2020MNRAS.492.5234Z}
{Zdziarski} A.~A.,  {Szanecki} M.,  {Poutanen} J.,  {Gierli{\'n}ski} M.,
  {Biernacki} P.,  2020, \mn@doi [\mnras] {10.1093/mnras/staa159}, \href
  {https://ui.adsabs.harvard.edu/abs/2020MNRAS.492.5234Z} {492, 5234}

\bibitem[\protect\citeauthoryear{{Zhang} et~al.,}{{Zhang}
  et~al.}{2020}]{2020MNRAS.494.1375Z}
{Zhang} L.,  et~al., 2020, \mn@doi [\mnras] {10.1093/mnras/staa797}, \href
  {https://ui.adsabs.harvard.edu/abs/2020MNRAS.494.1375Z} {494, 1375}

\bibitem[\protect\citeauthoryear{{{\.Z}ycki}, {Done}  \& {Smith}}{{{\.Z}ycki}
  et~al.}{1999}]{1999MNRAS.309..561Z}
{{\.Z}ycki} P.~T.,  {Done} C.,   {Smith} D.~A.,  1999, \mn@doi [\mnras]
  {10.1046/j.1365-8711.1999.02885.x}, \href
  {https://ui.adsabs.harvard.edu/abs/1999MNRAS.309..561Z} {309, 561}

\makeatother
\end{thebibliography}







\bsp	
\label{lastpage}
\end{document}